\documentclass[letterpaper,preprint,amsmath,amsfonts,amssymb,showpacs]{revtex4}
\usepackage{graphicx}
\newcommand{\RR}{\mathbb{R}}
\newcommand{\ZZ}{\mathbb{Z}}
\newcommand{\R}{{\bf R}}

\newcommand{\y}{{\bf y}}
\newcommand{\avec}{{\bf a}}
\newcommand{\Avec}{{\bf A}}
\newcommand{\bvec}{{\bf b}}
\newcommand{\Bvec}{{\bf B}}
\newcommand{\cvec}{{\bf c}}

\newcommand{\tvec}{{\bf t}}
\newcommand{\rvec}{{\bf r}}
\newcommand{\wvec}{{\bf w}}

\newcommand{\beq}{\begin{equation}}
\newcommand{\eeq}{\end{equation}}
\newcommand{\beqs}{\begin{eqnarray}}
\newcommand{\eeqs}{\end{eqnarray}}

\newcommand{\evec}{{\bf e}}
\newcommand{\fvec}{{\bf f}}

\newcommand{\mvec}{{\bf m}}

\newcommand{\uvec}{{\bf u}}

\newcommand{\xvec}{{\bf x}}
\newcommand{\yvec}{{\bf y}}
\newcommand{\zvec}{{\bf z}}
\newcommand{\Fvec}{{\bf F}}
\newcommand{\Gvec}{{\bf G}}
\newcommand{\Hvec}{{\bf H}}
\newcommand{\Ivec}{{\bf I}}
\newcommand{\Mvec}{{\bf M}}
\newcommand{\Qvec}{{\bf Q}}
\newcommand{\Rvec}{{\bf R}}
\newcommand{\Svec}{{\bf S}}

\newcommand{\Uvec}{{\bf U}}
\newcommand{\Vvec}{{\bf V}}
\newcommand{\Wvec}{{\bf W}}

\newcommand{\calB}{{\cal B}}
\newcommand{\calL}{{\cal L}}
\newcommand{\vphi}{{\varphi}}
\newcommand{\eps}{{\varepsilon}}

\begin{document}

\title{An elasticity theory for self-assembled protein lattices
with application to the martensitic phase transition in bacteriophage T4 tail sheath}
\author{Wayne Falk}
\email[]{falk@aem.umn.edu}
\affiliation{Department of Diagnostic and Biological Sciences, University of Minnesota, Minneapolis, MN 55455}
\author{Richard D. James}
\email[]{james@umn.edu}
\affiliation{Department of Aerospace Engineering and Mechanics, University of Minnesota, Minneapolis, MN 55455}

\date{\today}

\begin{abstract}
We propose an elasticity theory for one and two dimensional
arrays of globular proteins
for which the free energy is affected by relative
position and relative rotation between neighboring molecules.
The kinematics
of such assemblies is described, the conditions
of compatibility are found, a form
of the free energy is given, and
formulas for applied forces and moments are developed.
It is shown that fully relaxed states of sheets consist of
helically deformed sheets which themselves are composed
of helical chains of molecules in rational directions.

We apply the theory to
the fascinating contractile deformation that occurs in the
tail sheath of the virus bacteriophage T4, which aids its
invasion of its bacterial host.  Using electron density
maps of extended and contracted sheath, we approximate
the domains of each molecule by ellipsoids and then evaluate
our formulas for the position and orientation of each molecule.
We show that, with the resulting kinematic description, the
configurations of extended and contracted tail sheath are
generated by a simple formula. We proposed
a constrained version of the theory based on measurements
on extended and contracted sheath.
Following a suggestion
of Pauling [{\em Discussions of the Faraday Society} {\bf 13}, 170-6 (1953)], 
we develop a simple model of
the molecular interaction.   The resulting free energy
is found to have a double-well structure.
Certain simple deformations are studied (tension, torsion
inflation); the theory predicts a first-order Poynting effect
and some unexpected relations among moduli.
Finally, the force of penetration is given, and
a possibly interesting program of epitaxial
growth and patterning of such sheets is suggested.

\end{abstract}

\pacs{87.10.+e}

\maketitle 

\tableofcontents

\section{Introduction}   \label{sect1}
A remarkably large number of biological structures are composed
of identical protein molecules, or mixtures
of a few different protein molecules, in regular arrays.  Examples
are microtubules, bacterial flagella, F-actin filaments
and viral coats.  These are different
from typical inorganic crystals in that the individual molecules
are composed of many atoms and the whole array is typically
not a 3-D crystal but is often a single molecule thick regular
array on a sheet, either flat, curved or polyhedral, or else
in a linear chain.    Often the latter adopt helical forms,
and, in the example below of bacteriophage T4 tail sheath,
the cylindrical sheath is composed of two families of helical
chains of proteins.
Besides the reduced dimensionality and natural curvature, the
protein-protein interactions involve one or more bonding sites
with groups of bonded atoms distributed over the bonding site.
Because of this, the interactions
can be complex (from a first principles' viewpoint) and proteins
exert both forces and moments on each other.  However, a simplifying
feature of the interactions is that individual protein molecules
in such arrays predominantly interact only with nearest neighbors.

With the rapid development of optical tweezers and atomic
force microscopes \cite{smith},\cite{ivanoska}, it
has become possible to subject a protein structure to a
force or moment and to measure its elastic response.
These experiments seem to be often
interpreted in terms of classical macroscopic
theories of elasticity.  For example, Kirchhoff's rod theory
is often used
to interpret experiments on chains such as DNA \cite{bustamante},
\cite{maddocks1}, \cite{maddocks2}.   As discussed by these
authors, Kirchhoff's rod theory is expected to be valid
when the length of the chain is much larger than its
radius of curvature, and it has been used successfully
in such cases.
But Kirchhoff's rod theory
is built on certain assumptions relating to the macroscopic theory
of nonlinear elasticity.  Thus, for example, in its simplest
anisotropic form, three suitable bending-torsion experiments
suffice to determine the moduli, meaning that the mechanical behavior
in all subsequent experiments is then determined.
In addition, with only a few molecules (138 in the case of
T4 tail sheath) or with localized large curvatures, a
molecular elasticity theory may be needed.

For these reasons we develop here an elasticity theory that is
suitable for these protein arrays.  The proteins themselves
have irregular yet well-defined shapes and they interact
via localized bonding sites.  We steer a course midway
between detailed first principles calculations on the given protein
sequence (which, at present, would leave us stuck on the protein folding
problem) and macroscopic nonlinear elasticity.  First, each molecule is
given a position and orientation.  We explain how position and
orientation are related to
detailed structure, in a manner that is consistent with the results
of first principles calculations.   Our definition is
different from the usual one, but seems to have some
advantages.  It also has the property that the position
and orientation are given by a translation and rigid
rotation of a reference molecule that best approximates, in a least
squares sense, the deformed molecule.
As for interactions, we focus on pairwise forces and moments,
because bonding between globular proteins is often localized
at the region of contact between two proteins.  Because
we have an orientation variable as well as a positional one, our
theory also has points of contact with the theory of liquid
crystals, but  in
the end the theory is quite different.

We propose a form of the free energy based on these ideas
(Section \ref{sect3}).
Imposing the condition of frame-indifference, we define
certain ``strain variables'' upon which the free energy
depends.  Conditions of compatibility concern the extent
to which these strain variables can be assigned independently:
for sheets, we find that necessary and sufficient
conditions for compatibility (of a simply connected
sheet, defined precisely below) are that a certain pair of
sums of four terms vanishes.  These relate to the process of
checking compatibility around elementary squares consisting
of four molecules.  We find these conditions to be very
useful.

\begin{figure}
\includegraphics[width=.9\columnwidth]{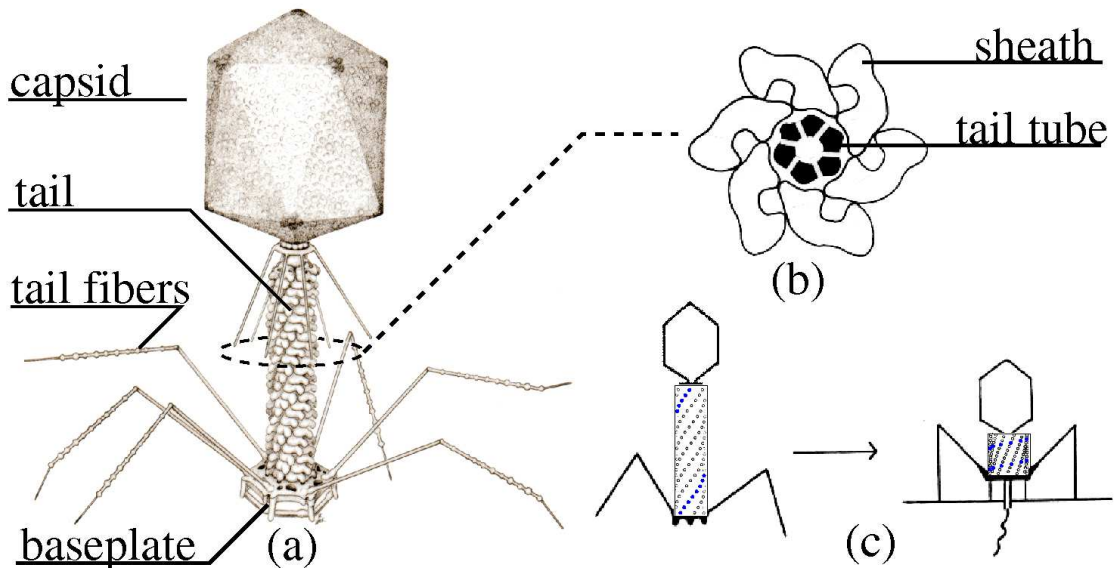}
\caption{(a) Structure of bacteriophage T4, based on
electron microscope
structure analysis to a resolution of about 2 to 3 nm,
from \cite{eiserling}. Reproduced with permission
of Fred Eiserling. (b) A cross section
showing concentric tail tube and sheath annuli, (c)
Schematic of contraction process \cite{O1982}.}
\label{figure1}
\end{figure}

For both chains and sheets, helical configurations arise as
the natural ordered structures.  In particular, under
general conditions
they are the free energy minimizers in the absence
of boundary conditions.
These results can be considered as analytical expressions of the
ideas of Crane \cite{crane}.
Ultimately the reason is the same reason that
these configurations are
natural in nonlinear elasticity (Ericksen \cite{ericksen}).
To oversimplify, in a helical configuration
any two parts of the chain, of equal monomer length,
are related by a Galilean transformation.  Therefore,
the equilibrium
of a short section implies, via rotation and translation,
that the whole helix is in equilibrium. In the present
context they play a deeper role: fully relaxed states of
a sheet governed by our free energy consist of helically
deformed sheets which themselves are composed of helical
chains of molecules (Section \ref{sect9}).

Using this theory, we study of deformations of the tail
sheath of bacteriophage T4, Figure \ref{figure1}.
Bacteriophages are viruses
that attack bacteria.   The T4 virus is
composed of a \emph{capsid}
containing the viral DNA (Figure \ref{figure1}a)
and a \emph{tail} shown extending down from the capsid.
 The tail consists of
a pair of concentric cylinders (Figure \ref{figure1}b)
each
about 1000 \AA\ long.  The inner \emph{tail tube} with a
diameter of
about 90 \AA\ is surrounded by the tail
sheath with an outer diameter
of about 240 \AA.  The sheath is composed of six parallel
helices, made
from chains of a single type of protein.  Although
the tail sheath is single molecular layer protein sheet, it
should not be considered (as is assumed by nonlinear plate
theory) thin relative to its radius of curvature, this ratio
being (thickness/mean radius of curvature) $\approx 1$.

Prior to invasion of the host, the sheath proteins are
arranged as steeply pitched
helices, and the tail adopts so-called {\it extended} structure.
During the virus' attack on a bacterium, the tail sheath changes
shape dramatically; the protein
helices compress, causing the sheath to shorten and fatten into a
more compact {\it contracted}
structure. This drives the relatively rigid inner
tail tube through the cell wall,
making a passage for the viral
DNA to pass into the host (Figure \ref{figure1}c).  During this process
the sheath contracts irreversibly to about 1/3 of its
original length accompanied by a 50\% increase in outer diameter.
The transformation has many features in common with martensitic
phase transformations, as has been noted by Olson and Hartman
\cite{O1982}.
For a general review of T4 tail structure and function see
Coombs and Arisaka.~\cite{coombs1994}

Once the viral DNA is inside the host, it reprograms the host
cell to produce all ingredients needed to form new viruses.
The viral protein molecules produced by this process self-assemble
into virus progeny  within the host, eventually causing it to burst,
releasing the viruses to infect other hosts.

We apply the theory of the protein sheet to the tail sheath of
bacteriophage T4. Using measured electron density maps of
extended and contracted sheath \cite{leiman}, \cite{leiman1},
we identify three domains
in each molecule and approximate these by ellipsoids.  We
then define the structures of the two phases.  We develop a
simple formula that produces these structures (\ref{T4sheath}),
and applies to any any model of the molecule, however complex.
The simplicity of this formula arises from the fact that,
even though the molecule itself may be complex, the relationship
between different molecules is very simple.
Based on the experimentally observed mode of
deformation, we adopt a constrained
theory for the sheath.  The constraints are exactly satisfied
by both extended and contracted sheath.  The resulting theory
has surprising implications with regard to the response of
the sheath to different loadings, including a strong first-order
Poynting effect, unexpected relations between moduli, certain
combinations of applied axial force and moment that do no
work on the sheath, and a certain relation between the force and
moment needed to transform contracted to extended sheath.
Among these
results, there are numerous possible points of comparison
with future small-scale quantitative experiments.

A very early model of helical contraction proposed by Pauling
\cite{pauling} provides a basis for simplifying our free
energy.  Pauling envisioned that helices forming a
cylinder could be compressed
to the point where adjacent turns of the helix would form
bonds.  This leads to a simple model of interactions with
one family of bonds guiding assembly and another causing
contraction.  As higher resolution images of T4 tail sheath
become available \cite{leiman}, this remains likely the principal
mechanism for sheath contraction.  Guided by Pauling's
mechanism, we build a simplified free energy for T4 tail
sheath.

The deformation of T4 tail sheath is a particularly interesting
case for the theory for several reasons: 1) very large changes
of shape in an organized protein structure take place as a
means of producing force, 2) the shape change has been identified as
of martensitic type, which suggests a multi-well elasticity
theory, and 3) there is an interest in understanding how this
phase transformation relates to nonbiological martensitic transformations,
and in particular how the force and energy of contraction compare
in the biological and nonbiological cases.  The latter could
suggest strategies for man-made analogs of the T4 tail sheath.
Finally, 4) a quantitative evaluation of the energy stored
has interesting biological significance.  That is, in bacteriophage T4,
as in all viruses, there is no mechanism for the production of
energy.  Thus, all the energy that is released upon contraction
of the tail sheath
must be stored during the assembly phase of the virus, from
apparently high free energy molecules created during translation
of the viral genome, aided by the energy consuming translation
mechanism of the bacterial host.  For the purpose of storage of
this energy, the process of epitaxial stabilization, familiar
from the growth of semiconductor compounds on single crystal
substrates, apparently plays an
important role.  Motivated by these ideas we suggest such a
program of epitaxy (Section 10).

Mathematical Notation: bold faced uppercase letters are 3 $\times$ 3 matrices and
bold faced lowercase letters are vectors in $\RR^3$. Components are relative
to a fixed orthonormal basis throughout.  The summation convention
is used, $\Avec \cdot \Bvec$ is the
inner product between matrices: $\Avec \cdot \Bvec = A_{ij}B_{ij}$,
and $\parallel \! \Avec \! \parallel = \sqrt{\Avec^T \Avec}$.
The superscript T denotes transpose, tr$\Avec = A_{ii}$ is the trace of $\Avec$, $\Ivec$
is the identity matrix,
$(\avec \otimes \bvec)$ is the matrix with components $a_i b_j$,
Skew$\Avec = (1/2)(\Avec - \Avec^T)$,
and $3 \times 3$ rotation matrices are denoted by
SO(3) $= \{\Rvec : \Rvec^T \Rvec = \Ivec, {\rm det}\ \Rvec = +1 \}$.
$\ZZ^2$ denotes all pairs of integers $(i,j)$.
$\eps_{ijk}$ is the permutation symbol, defined by
$\eps_{ijk} = 1$ if $ijk$ is an even permutation of $123$,
$\eps_{ijk} = -1$ if $ijk$ is an odd permutation of $123$,
and $\eps_{ijk} = 0$ if any index is repeated.

\section{Kinematic description of individual protein molecules, chains
and sheets}   \label{sect2}

We are interested in chains and sheets consisting of protein
molecules.  For simplicity we shall consider both structures to
consist of identical molecules\footnote{The extension to regular arrays
of several different proteins is expected to be similar.}.

A molecule will be specified by
a pair $(\yvec, \Rvec)$ consisting of a position vector
$\yvec \in \RR^3$ and a rotation matrix $\Rvec \in$ SO(3),
termed, respectively, position and orientation.
Protein chains and sheets are constructed by building up one and
two dimensional arrays of these molecules.  For chains we choose
a set of integers $\{1 , \dots, N  \}$ corresponding to N
molecules and assign mappings
\beq
 \yvec: \{1 , \dots, N  \}  \to \RR^3, \ \ \ \
 \Rvec: \{1 , \dots, N  \}  \to {\rm SO(3)},
\eeq
For sheets we denote molecules by pairs of integers $(i,j) \in \ZZ^2$.
We consider a set ${\cal D} = \{1, \dots, N\} \times \{1, \dots, M\}$
and mappings
\beq
 \yvec: {\cal D}  \to \RR^3, \ \ \ \
 \Rvec: {\cal D}  \to {\rm SO(3)},
\eeq
For the configuration of a single molecule
we use the obvious notation $(\yvec_i,\Rvec_i)$ for chains and
$(\yvec_{i,j},\Rvec_{i,j})$  for sheets.  More generally,
${\cal D}$ could have the form $\Omega \times \ZZ^2$ where
$\Omega$ is a domain in the plane.

The usual way to define orientation of a molecule, or a collection
of molecules, is to take moments of the (time averaged)
mass distribution (see, e.g., De Gennes, \cite{degennes},
Chapter 2), typically the second or fourth moment.
Below, we suggest a different definition of orientation that
seems to have advantages with regard to the connection with
molecular dynamics simulation.

Each molecule consists $\nu$ atoms of C, H and various
other elements in a folded configuration.
For each atom we assign a corresponding atomic mass $m_i\, ,
i = 1, \dots, \nu $.  Near physiological temperatures, the atoms
in a protein molecule undergo rather large vibrations, but it is
still sensible to talk about time averaged position on macroscopic
time scales and we shall assign these average positions as
$\yvec_1, \dots, \yvec_{\nu}$.  Time averaged momentum
has dynamical
significance, so we use it to define the position of
a molecule.  We will take its {\it position} to be its mass averaged
position,
\beq
\yvec = \frac{\Sigma_{i = 1}^{\nu} m_i \, \yvec_i}{\Sigma_{i = 1}^{\nu} m_i }.
\label{ypos}
\eeq
From the same information we can also obtain a measure of the orientation
of a molecule.  That
is, we consider a molecule in standard position defined by fixed
atomic positions $\xvec_1, \dots, \xvec_{\nu}$. This standard
position could for example be the collection of
positions in a crystallized
form of the molecule, deduced from x-ray crystallography, or
from theoretical studies of the configuration of
single molecules in solution.   From these reference positions
we define the mass averaged reference position as above,
\beq
\xvec = \frac{\Sigma_{i = 1}^{\nu} m_i \, \xvec_i}
{\Sigma_{i = 1}^{\nu} m_i }\ .
\label{xpos}
\eeq
A natural concept of orientation is obtained through the
average deformation
gradient of the molecule, defined in the following way,
\beq
  \Fvec = \frac{\Sigma_{i = 1}^{\nu} m_i \, (\yvec_i - \yvec) \otimes
  (\xvec_i - \xvec)}{r^2 \sum_{i=1}^{\nu} m_i}.  \label{defgrad}
\eeq
Here, $r$ can be taken as a typical radius of the reference
molecule, e.g.,
\beq
r =\sqrt{\frac{\sum m_i (\xvec_i -\xvec)^2}{\sum m_i}}.
\eeq
(The position and orientation of a molecule will turn
out to be independent of $r$).
The expression  (\ref{defgrad}) for $\Fvec$
is dimensionless and translation invariant.  Typically, it
will be true that $\det \Fvec >0$, which we assume.
If the $\yvec_i$ represent a rigid deformation of the reference
molecule, i.e., $\yvec_i = \Qvec \xvec_i + \cvec, \
i = \{1, \dots, \nu \}, \  \Qvec \in$ SO(3),
then it follows from (\ref{defgrad}) that
\beq
  \Fvec = \Qvec \Vvec.     \label{pdq}
\eeq
where
\beq
  \Vvec = \frac{\Sigma_{i = 1}^{\nu} m_i \, (\xvec_i - \xvec) \otimes
  (\xvec_i - \xvec)}{r^2 \sum_{i=1}^{\nu} m_i}.    \label{pdu}
\eeq
The latter  is
interpretable as a normalized reference moment of inertia.
In general we will use the rotation
in the polar decomposition of $\Fvec$ as a measure of
orientation.  That is, we will write
\beq
\Fvec = \Rvec \Uvec\ {\rm where}\ \Rvec \in\ {\rm SO(3)}\
{\rm and}\ \Uvec = \Uvec^T\  {\rm is\ positive\!-\!definite},
\label{pd}
\eeq
and define $\Rvec$ in this decomposition as the {\it orientation}.
For molecules that deform as well as
rotate and translate, $\Rvec$ is still a natural measure of
orientation.

These definitions of position and orientation, the latter
defined by a polar
decomposition of the apparently
complicated formula for $\Fvec$, have several
attractive features.   First,
we observe that both $\Fvec$  and  $\yvec$ are linear functions of
the positions $\yvec_1, \dots \yvec_{\nu}$.  We have framed the
definitions in this way so that their second time derivatives
are immediately related to (time averaged) forces via the equations
of molecular dynamics.
Second, there is a useful variational characterization
of $\Rvec$  and  $\yvec$.  That is,
$\Rvec (\xvec_i -\xvec)+ \yvec$ is the rigid deformation that
best approximates the mass distribution of the
molecule in the least squares sense.
More precisely, if we consider
the rotation matrix $\Qvec$ and vector $\cvec$ that
minimize\footnote{We write this as an integral rather
than a sum to indicate that this characterization of
$\Rvec$ and $\yvec$ applies to
any mass distribution.},
\beq
    \min_{\cvec\, ,\Qvec \in {\rm SO(3)}} \int \big|\, \yvec(\zvec)
    - [\Qvec(\zvec-\xvec) + \cvec)]\,  \big|^2 dm(\zvec),   \label{mindef}
\eeq
where $dm$ is the mass measure of the molecule,
i.e., $m = \sum m_i \delta_{x_i}$, and $\yvec(\xvec_i) = \yvec_i$,
then it follows from the simple quadratic minimization problem
(\ref{mindef}) that
$\Qvec = \Rvec$ and $\cvec = \yvec$ as defined by
(\ref{ypos}) and (\ref{defgrad}),(\ref{pd}).
The proof of this fact is straightforward.  First, by differentiating
(\ref{mindef}) with respect to $\cvec$ we conclude immediately that
$\cvec = \yvec$.  Then we replace $\cvec = \yvec$ in (\ref{mindef})
and simplify. The minimization over $\Qvec \in $ SO(3) then becomes
\beqs
\min_{\Qvec \in {\rm SO(3)}} (-{\rm tr} (\Qvec^T \Fvec))
&=& -\max_{\Qvec \in {\rm SO(3)}}{\rm tr}(\Qvec^T \Fvec) =
-\max_{\Qvec \in {\rm SO(3)}}{\rm tr}(\Qvec^T \Rvec \Uvec) \nonumber  \\
&=&
-\max_{\bar{\Qvec} \in {\rm SO(3)}}{\rm tr}(\bar{\Qvec} \Uvec)
= -\max_{\bar{\Qvec} \in {\rm SO(3)}}\sum_i \lambda_i \evec_i \cdot
\bar{\Qvec} \evec_i,
\eeqs
 where $\{\evec_1, \evec_2, \evec_3\}$ are
orthonormal eigenvectors of $\Uvec$ with corresponding positive eigenvalues
$\{\lambda_1, \lambda_2, \lambda_3\}$.  It is immediately seen that
the latter maximization problem is uniquely solved by $\bar{\Qvec} = \Ivec$,
implying that $\Qvec = \Rvec$.

The orientation $\Rvec$ can be obtained from the
formula
$\Rvec = \Fvec \Uvec^{-1} = \Fvec (\sqrt{\Fvec^T \Fvec})^{-1}$,
the square root being the unique positive-definite square root.
For the purpose of a constrained molecular dynamic simulation
(with given $\Rvec$) it is useful to have a linear constraint.
A necessary\footnote{This condition is
nearly sufficient for the determination of $\Rvec$, the only
freedom being that if $\Rvec$ satisfies (\ref{skewcond})
then so does $\Rvec \hat{\Rvec}$ where $\hat{\Rvec}$ is a
180$^\circ$ rotation about one of the eigenvectors of
$\Uvec$.  In practice, if the configuration is changing
slowly, this nonuniqueness would not cause a problem.}
condition that $\Rvec$ is related to $\Fvec$ by
(\ref{pd}) is that Skew$(\Rvec \Fvec^T) = 0$, that is,
\beq
{\rm Skew} \left( \Rvec \left( \Sigma_{i = 1}^{\nu} m_i \,
  (\xvec_i - \xvec)\otimes (\yvec_i - \yvec) \right) \right) = 0,
  \label{skewcond}
\eeq
which is a linear constraint on the $\yvec_i$.
 

In summary, our basic kinematics of a molecule is specified by
a pair $(\yvec, \Rvec)$ defined by (\ref{ypos}) and
(\ref{defgrad}),(\ref{pdu}),(\ref{pd}).
We wish to emphasize again that this choice of kinematics
does not entail assumptions of rigidity of molecules.  These
formulas do allow one to determine
our kinematic variables from unrestricted
first principles calculations.

We note that there are more complicated theories possible
with this general type of kinematics.
A further possible generalization
would be that the free energy is affected by position and
and full deformation gradient of molecules defined
by (\ref{defgrad}).

\section{Free energy}   \label{sect3}

Although all atoms have infinite range,
each protein molecule interacts primarily with its close neighbors
and we shall develop the theory on this basis.  Because bonding
sites are often localized and interlocking,
molecules are expected to exert
both forces and moments on neighboring molecules.

We base the theory on a formula for the free energy.  We make
two simplifying assumptions which easily could be
generalized: 1) the molecules are
identical (Thus we can use the reference configuration
introduced in Section \ref{sect2} for all molecules), and
2) we only consider nearest neighbor interactions.  For chains,
the ``nearest neighbor'' of $i \in \{ 2, \dots, N-1 \}$
refers to the two molecules $i-1, i+1$ for interior molecules
while it refers to the molecule $2$ for $i = 1$ and
$N-1$ for $i = N$.  For sheets there are various possibilities.
One can have triangular lattices with each molecule bonded
to 6 nearest neighbors or rectangular lattices with each molecule
having 4 nearest neighbors or more complicated situations.
In the case of 4 nearest neighbors
the nearest neighbors of $(i,j)$ consist of all
molecules of the form $(i+1,j),(i-1,j),(i,j+1),(i,j-1)$
that lie in ${\cal D}$.  If not all four of these are in
${\cal D}$
we call $(i,j)$ a {\it boundary molecule}; otherwise, we call
it an {\it interior molecule}.  Here we  write the
free energy only in the case of 4 nearest neighbors,
the generalizations being automatic.

Since such protein arrays are of interest in solution, the
assumptions are somewhat different than would be appropriate
for atoms in a polymeric chain or a crystal. In particular
the ``free energy'' will be taken as the  free energy
of the protein assembly and a fixed  volume V of
the surrounding solution.   This is appropriate to the
case that V is surrounded by a large bath B
having fixed temperature and fixed
chemical potentials of species in solution.
All free energies below will
depend on the
temperature and chemical potentials,
but since these will be fixed throughout this paper,
we will leave these parameters
out of the notation.
As is well-known the presence of the solution profoundly
affects the free energy of the protein through osmotic
effects, but it also affects the form of the free energy.
In particular, boundary molecules may have a free energy that
is different from interior molecules, because one or more
of their bonding sites is unbonded and exposed directly to
the solution.

In this simplest situation we will assume that, for chains,
there is a molecular interaction free energy which depends
on the position and orientation of a pair of molecules
$\psi(\avec, \Rvec, \bvec, \Qvec)$
defined for $\avec, \bvec \in \RR^3$ and $\Rvec, \Qvec
\in$ SO(3).  In order to accommodate the possibility of boundary 
effects, we distinguish the free energy contribution from
the interaction of the first and second molecules,
$\psi_1(\avec, \Rvec, \bvec, \Qvec)$  and the next to
last and last ones,
$\psi_N(\avec, \Rvec, \bvec, \Qvec)$.  The total free energy is
then,
\beqs
\Psi(\y_1,\Rvec_1\, , \ldots,\, \y_N, \R_N)
&=& \sum_{i=2}^{N-2} \psi(\y_{i},\R_{i},\y_{i+1},\R_{i+1})  \\ \nonumber
&+& \psi_1(\y_{1},\R_{1},\y_{2},\R_{2}) +
\psi_N(\y_{N-1},\R_{N-1},\y_{N},\R_{N})\label{Psichain}
\eeqs
The pairwise form of this free energy is justified
by the presence of localized bonding sites.
While it would seem like the molecular
free energies $\psi_1$  and $\psi_N$ could be quite
different from each other and from $\psi$, in fact this is
not true within the present context.  To see this, consider
a chain $\{1, \dots, M, M+1, \dots, N\}$ and translate uniformly
the molecules $\{M+1, \dots, N\}$.  Physically, if this translation
is large, this should give the sum of the energies of two
separate chains.  If one writes this out for an arbitrary
configuration, one finds that, in
fact, $\psi_1 = \psi_N = \psi$, so the preceding energy is
in fact,
\beq
= \sum_{i=1}^{N-1} \psi(\y_{i},\R_{i},\y_{i+1},\R_{i+1}).
\label{Psichain1}
\eeq
The main physical assumption embodied here is
that the contribution to the free energy from a pair of
molecules is unaffected by the positions and orientations
of all other molecules in the chain, in keeping with the
idea that the main free energy changes are due to changes
in conformation at the bonding site between a pair of molecules.

For sheets the assumptions are analogous.  In this case there
are 15 different kinds of boundary molecules, depending on
which of the four bonds is missing.  To simplify the notation
we let $\calB$ be the set of boundary molecules and write
the total free energy as
\beqs
\lefteqn{\Psi(\y_{1,1},\Rvec_{1,1}\, , \ldots,\, \y_{N, M}, \R_{N, M})}
 \nonumber  \\
&=& \sum_{(i,j)\, \in\, \ZZ^2\, \cap\, {\cal D}\setminus \calB }
\psi_1(\y_{i,j},\R_{i,j},\y_{i+1,j},\R_{i+1,j}) +
\psi_2(\y_{i,j},\R_{i,j} ,\y_{i,j+1},\R_{i,j+1}) \nonumber  \\
& & \ +  \sum_{(i,j)\, \in\, \calB} \psi^{\cal B}_{i,j},
\label{Psisheet}
\eeqs
where $\psi^{\cal B}_{i,j}$ is a free energy for the
boundary molecules.   Each $\psi^{\cal B}_{i,j}$ is one
of the 15 functions describing the free energy of interaction of
boundary molecules with dependence on the position and
orientation of neighbors that are present.  In writing this free
energy, we have effectively assumed that all bonding sites of
between molecules $i,j$ and $i, j+1$ are the same, independent
of $i$ and $j$, and the same for sites of the form $i, j$  and $i+1,j$.
Also, in many interesting cases not all of the 15 kinds of
boundary molecules are represented.
For example, in isolated T4 tail sheath as usually pictured,
there at only two kinds of boundary molecules.

The condition of frame-indifference restricts the form of
the molecular free energies.  We note first that, according
to the definitions (\ref{ypos}) and (\ref{pd}),
the quantities $\yvec_1$ and $\Rvec_1$ are transformed into
$\Rvec \yvec_1 + \cvec$ and $\Rvec \Rvec_1$ under a superimposed
rigid body motion $\Rvec \yvec_i + \cvec$ of all the atoms.
Thus, in the case of chains, the condition of frame-indifference
is (for interior molecules)
\beq
 \psi_(\yvec_1, \Rvec_1, \yvec_2, \Rvec_2) =
  \psi(\Rvec\yvec_1 + \cvec, \Rvec\Rvec_1, \Rvec\yvec_2
   + \cvec, \Rvec\Rvec_2),
\eeq
which must hold independently for $\Rvec, \Rvec_1, \Rvec_2 \in$ SO(3),
and $\cvec, \yvec_1, \yvec_2 \in \RR^3$.  Making the special
choice $\Rvec = \Rvec_1^T$ and $\cvec = -\Rvec_1^T \yvec_1$, we
see  that
\beq
 \psi(\yvec_1, \Rvec_1, \yvec_2, \Rvec_2) =
  \psi(0, \Ivec, \Rvec_1^T(\yvec_2 - \yvec_1),
  \Rvec_1^T\Rvec_2)  = \vphi(\tvec, \Qvec),
  \label{cfe}
\eeq
where $\Qvec = \Rvec_1^T \Rvec_2$  is a {\it relative orientation}
that is unaffected by rigid body rotations and
$\tvec = \Rvec_1^T (\yvec_2 - \yvec_1)$  is a {\it relative
translation} which is also unaffected by superimposed rigid
motions.  Note that $\Qvec$ is not simply the
 rotation which maps the orientation of
 molecule 2 into that of molecule 1 (or
{\it vice versa}), and
similarly, $\tvec$ is not a simple translation of 2 into
1.  Rather, these quantities behave like strains, and describe
the 6 degrees of freedom associated with the
straining of bond sites caused by changes of orientation
and relative position of 1 and 2.

    For sheets the restrictions of frame-indifference are
similar: the functions $\psi_1$ and $\psi_2$ in  (\ref{Psisheet})
both satisfy (\ref{cfe}).  We let $\vphi_{1,2}$ be the corresponding
reduced functions defined by (\ref{cfe}) subscripted by 1,2.

\section{Compatibility} \label{sect4}

Compatibility concerns the extent to which one can prescribe
the quantities describing strain, or, more generally, the
functions on which the free energy depends after the condition
of frame-indifference has been imposed.  In our case this concerns
the extent to which we can assign the relative translations and
relative orientations.

In the case of chains we therefore assign
$(\tvec_1,\Qvec_1, \dots, \tvec_{N-1}, \Qvec_{N-1})$
and ask whether there are positions and orientations
$(\yvec_1,\Rvec_1, \dots, \yvec_{N}, \Rvec_{N})$ consistent
with these in the sense that
$\tvec_i = \Rvec_i^T(\yvec_{i+1} - \yvec_i)$ and
$\Qvec_i = \Rvec_i^T \Rvec_{i+1}$, $i = 1, \dots, N-1$.
It is immediately seen
that these conditions are solvable, and all solutions are related to
each other by exact rigid motions of the entire molecule.
Hence, the problem of compatibility for chains is
analogous to the case of 1-D rod theories in continuum mechanics:
there are no conditions of compatibility and the freedom
is precisely overall rigid deformations.

For sheets there are nontrivial restrictions of compatibility, as
expected based on continuum shell theories.  We begin
by considering the case of 4 bonding directions.
For the analysis
below a {\it path} is a succession of nearest neighbors in
${\cal D}$ and a {\it loop} is a closed path.   We assume here that
${\cal D}$ is {\it discretely simply connected} in
the sense that every point is connected to every other point
by a path and any non-self-intersecting closed
loop in ${\cal D} \cap \ZZ^2$ is the boundary of
the union of (closed) unit squares contained in ${\cal D}$.  We assign
relative translations and relative orientations
\beq
(\tvec_{i,j}, \Qvec_{i,j}),   \ (i,j)\in {\cal D}^r \ \
 {\rm and} \ \ (\hat{\tvec}_{i,j},
\hat{\Qvec}_{i,j}), \  (i,j) \in {\cal D}^u,
  \label{tQ}
\eeq
where ${\cal D}^{r, u}$ are the subsets $(i,j) \in {\cal D}$ such that
$(i+1, j)$ (resp. $(i,j+1)$) are also in ${\cal D}$ (r denotes ``right''
and u denotes ``up'').  We ask whether there
are positions and orientations $\yvec_{i,j}, \Rvec_{i,j},
\ (i,j) \in {\cal D}$, that satisfy
\beqs   \label{tyQR}
\tvec_{i,j} &=&  \Rvec_{i,j}^T(\yvec_{i+1,j} - \yvec_{i,j})\, ,  \nonumber  \\
\Qvec_{i,j} &=&  \Rvec_{i,j}^T\Rvec_{i+1,j}\, , \quad \quad
 \quad \quad \quad (i,j) \in {\cal D}^r, \nonumber  \\
\hat{\tvec}_{i,j} &=&  \Rvec_{i,j}^T(\yvec_{i,j+1} - \yvec_{i,j})\, ,  \nonumber  \\
\hat{\Qvec}_{i,j} &=&  \Rvec_{i,j}^T\Rvec_{i,j+1}\, , \quad \quad
 \quad \quad \quad
(i,j) \in {\cal D}^u.   \label{obj_vars}
\eeqs
Immediately we see that there are some restrictions.  For example,
if $\Rvec_{i,j}$ has been determined, then by successive application
of (\ref{tyQR})$_{2,4}$
there are overdetermined equations for, say, $\Rvec_{i+1,j+1}$, these
being,
\beqs
\Rvec_{i+1,j+1} &=& \Rvec_{i,j+1} \Qvec_{i,j+1} =
\Rvec_{i,j}\hat{\Qvec}_{i,j} \Qvec_{i,j+1}  \nonumber  \\
\Rvec_{i+1,j+1} &=& \Rvec_{i+1,j} \hat{\Qvec}_{i+1,j} =
\Rvec_{i,j} \Qvec_{i,j} \hat{\Qvec}_{i+1,j}.
\eeqs
Equating these, we get
\beq
\hat{\Qvec}_{i,j} \Qvec_{i,j+1} =  \Qvec_{i,j} \hat{\Qvec}_{i+1,j},
\eeq
or,
\beq
\hat{\Qvec}_{i,j} \Qvec_{i,j+1}\hat{\Qvec}^T_{i+1,j}\Qvec^T_{i,j} = \Ivec,
\eeq
This has the following interpretation: as we go say clockwise
around a unit square in the lattice $\ZZ^2$, the product
of the $\Qvec$'s (taken with transpose if the path goes to the
left or down) is the identity.  Two neighboring squares, both
traversed clockwise, give such identities of the form
$\hat{\Qvec}_1 \Qvec_2 \hat{\Qvec}^T_3 {\Qvec}^T_1 = \Ivec$ and
$\hat{\Qvec}_3 \Qvec_4 \hat{\Qvec}^T_5 {\Qvec}_3^T = \Ivec$ which immediately
gives $\Ivec =  {\Qvec}^T_1 \hat{\Qvec}_1 \Qvec_2 \hat{\Qvec}^T_3
\hat{\Qvec}_3 \Qvec_4 \hat{\Qvec}^T_5 {\Qvec}_3^T  =
{\Qvec}^T_1 \hat{\Qvec}_1 \Qvec_2 \Qvec_4 \hat{\Qvec}^T_5 {\Qvec}_3^T$,
that is,
$ \hat{\Qvec}_1 \Qvec_2 \Qvec_4 \hat{\Qvec}^T_5 {\Qvec}_3^T {\Qvec}^T_1
 = \Ivec $; this is a similar compatibility
 condition for the rectangle consisting of the
 union of the two squares.
 By induction and using the discrete simple
 connectedness\footnote{By the discrete simple connectedness of
 ${\cal D}$, a non-self-intersecting loop in ${\cal D} \cap \ZZ^2$
 encloses a union of squares that can be completely exhausted
 by adding successive squares that share an edge. }
 of ${\cal D}$, this extends to any
 non-self-intersecting closed loop in ${\cal D} \cap \ZZ^2$.
 
So far, the argument concerns the solution of the last two equations
of (\ref{tyQR}).  For the translations, by again traversing
a unit square in the clockwise sense, we have from (\ref{tyQR})$_{1,3}$
that $\Rvec_{i,j} \hat{\tvec}_{i,j} + \Rvec_{i,j+1} {\tvec}_{i,j+1}
 - \Rvec_{i+1,j} \hat{\tvec}_{i+1,j} - \Rvec_{i,j} \tvec_{i,j} = 0$, which,
 after premultiplication by $\Rvec_{i,j}^T$ gives
\beq
 \hat{\tvec}_{i,j} + \hat{\Qvec}_{i,j} {\tvec}_{i,j+1}
 - \Qvec_{i,j} \hat{\tvec}_{i+1,j} -  \tvec_{i,j} = 0.
\eeq
As above, equations of this form for neighboring squares can
be combined to an equation of compatibility for a rectangle and
then, by iteration, to a non-self-intersecting closed loop.

By this time it is clear that the pattern of argument is essentially
the same as that for differentials (i.e., this kind of argument
does not really use that the differentials are small, if only
nearest neighbors interactions are considered).
That is,  necessary and sufficient conditions for (\ref{tQ})
to be compatible are that the compatibility conditions for
unit squares in ${\cal D}$, i.e., all equations of the form
\beqs
\hat{\Qvec}_{i,j} \Qvec_{i,j+1}\hat{\Qvec}^T_{i+1,j}\Qvec^T_{i,j} &=& \Ivec,
\nonumber \\
 \hat{\tvec}_{i,j} + \hat{\Qvec}_{i,j} {\tvec}_{i,j+1}
 - \Qvec_{i,j} \hat{\tvec}_{i+1,j} -  \tvec_{i,j} &=&  0
 \label{com}
\eeqs
are satisfied.  The necessity of these
conditions has been proved above.  The sufficiency  follows
by giving arbitrarily $\yvec_{0,0} \in \RR^3, \Rvec_{0,0} \in$ SO(3),
assuming without loss of generality
that $(0,0)$ is in ${\cal D}$.  Then,
for any other $(k,m) \in {\cal D} \setminus \calB$, we consider
a path from $(0,0)$ to $(k,m)$.  Successive application of
(\ref{tyQR}) determines first $\Rvec_{k,m}$ and then $\yvec_{k,m}$,
and every such $\Rvec, \yvec$ along this path.  By a process
of exhaustion, i.e., construct a path which does not cross
itself or any other path to a point whose values
$\Rvec,\yvec$ have not been determined from a previously determined
one, we then determine all values $\Rvec_{i,j}, \yvec_{i,j}$.
These satisfy all of the equations  (\ref{tyQR}).  That is,
by construction, a point $(i,j)$ and neighbor $(i+1,j)$
(resp.,  $(i,j+1)$) are each connected to
$(0,0)$ by a path used in the construction.
These paths may coincide over
some initial length, but, once they depart from each other, they
never intersect.   Thus, by possibly shortening the loop,
we can  without loss of generality
assume the paths form a non-self-intersecting loop with a
single link removed.
Satisfaction of (\ref{tyQR}) then holds as
a consequence of
the compatibility condition for such loops.

There is clearly also uniqueness of the construction of
$\Rvec_{i,j}, \yvec_{i,j}$ up to the choice of
$\Rvec_{0,0}, \yvec_{0,0}$, which, by the frame-indifference
of the quantities  $\Qvec, \tvec$ is equivalent to uniqueness
up to overall rigid deformation.

Suppose now we add additional bonding directions.  Since
the equations (\ref{com}) are both necessary and sufficient
for the existence of the positions and orientations,
and these positions and orientations
are uniquely determined up to overall translation and
rotation (which does not affect the $(\tvec, \Qvec)$'s),
then all values of $(\tvec, \Qvec)$ corresponding to
other bonding directions are uniquely determined by
$(\tvec_{i,j}, \Qvec_{i,j}),   (\hat{\tvec}_{i,j},
\hat{\Qvec}_{i,j})$.  One can write formulas for these.
For example, if (as in T4 sheath) we have the additional
bonding directions
$(i,j)-(i-1,j+1)$, then
\beqs
 \bar{\tvec}_{i,j} &\stackrel{\rm def}{=}& \Rvec_{i,j}^T(\yvec_{i-1,j+1} - \yvec_{i,j})
  = \Qvec_{i-1,j}^T (\hat{\tvec}_{i-1,j} - {\tvec}_{i-1,j}),
 \nonumber \\
 \bar{\Qvec}_{i,j} &\stackrel{\rm def}{=}& \Rvec_{i,j}^T \Rvec_{i-1,j+1}
  = \Qvec_{i-1,j}^T \hat{\Qvec}_{i-1,j}.  \label{additional}
\eeqs
In summary, with additional bonding directions, necessary
and sufficient conditions for compatibility are
(\ref{com}), together with the formulas of the type
(\ref{additional}) that uniquely determine the values of
$(\tvec, \Qvec)$ for these additional directions in terms
of $(\tvec_{i,j}, \Qvec_{i,j}),   (\hat{\tvec}_{i,j},
\hat{\Qvec}_{i,j})$.

\section{Helical configurations} \label{sec5}
 
As shown in Figure \ref{figure4} below,
the tail sheath of bacteriophage T4 is
a sheet consisting of the union of two families of helices.
Helical structures arise often in biology and they also have
special position within the context of the present theory,
as we explain in this and the following sections.

From a purely geometric viewpoint H. R. Crane in 1950 \cite{crane}
argued that if two proteins have complementary bonding
sites and molecules bond at a specific angle,
then chains of these molecules are likely to from
helices.   In nonlinear elasticity
of rods and plates, helical configurations also arise in a
natural way (Chouaieb and Maddocks, Moakher, Maher and Maddocks,
\cite{maddocks1}, \cite{maddocks2}),
and in 3-D nonlinear elasticity
there are deep connections between the existence of helical
configurations in equilibrium, invariance and Noether's
theorem (Ericksen \cite{ericksen}).  In all
of these arguments the frame-indifference of the free
energy plays a central role, allowing the variables describing
strain to be either constant (in rod theory) or else to
depend on fewer reference coordinates for these special
configurations.

We begin with chains and recall from (\ref{cfe})
that the molecular free energy
depends on relative translation  and orientation
$(\tvec, \Qvec)$.  Motivated by the examples just cited
we first try to figure out
what are all configurations $(\y_1,\Rvec_1\, , \ldots,\, \y_N, \R_N)$
having constant values of $(\tvec, \Qvec)$.  In the following
section we explain the energetic significance of this choice,
beyond the obvious fact that such configurations have the property
that the molecular free energy is independent of the molecule.
This problem is immediately solved by the considerations of
compatibility of the preceding section; we have to solve
\beq
\Rvec_i^T (\yvec_{i+1} - \yvec_i) = \tvec,  \ \ \ \
\Rvec_i^T \Rvec_{i+1} = \Qvec, \ \ i = 1, \dots, N-1.
\label{helix0}
\eeq
and the general solution is,
\beq
\Rvec_{i+1} = \R_1\Qvec^{i}, \ \ \  \yvec_{i+1} = \yvec_1
 + \Rvec_1 \sum_{j = 0}^{i-1}\Qvec^{j} \tvec,  \ \ i = 1, \dots, N-1,
 \label{helix1}
\eeq
where $\yvec_1 \in \RR^3, \R_1 \in$ SO(3) are arbitrary,
and they are also the values of position and orientation
corresponding to molecule 1.   It is clear from
(\ref{helix1}) that the choice
$\yvec_1 \in \RR^3, \R_1 \in$ SO(3) also corresponds to
an arbitrary superimposed rigid deformation of the whole
array.

The positions of the molecules described by the
equations (\ref{helix1})
lie on a helix.  To see this put $\Rvec_1 = \Ivec$
and note that  $\Qvec = \Ivec$ corresponds to
the degenerate case of a molecules spaced equally along
a line all with the same orientation.  So we assume
 henceforth that $\Qvec \ne \Ivec$.  Then
$\Qvec$ has an axial vector, that is, a  vector
$\evec \in \RR^3$ whose direction is uniquely determined
such that $\Qvec \evec = \evec$.  By suitable choice of
the magnitude of $\evec$, we can
decompose $\tvec =  \tvec^{\parallel} +  \tvec^{\perp}$,
$\tvec^{\perp} \cdot \evec = 0,\ \tvec^{\parallel} \parallel \evec$;
then the second
of (\ref{helix1}) becomes
\beq
 \yvec_{i+1}  =  \yvec_1 +  i\, \tvec^{\parallel}  +
  \sum_{j = 0}^{i-1} \Qvec^{j}\tvec^{\perp}.
 \label{prohelix}
\eeq
The last term in (\ref{prohelix}) can be further
simplified.  To do so, note that $\Qvec - \Ivec$ is invertible
on the plane perpendicular to $\evec$ and define
$\rvec$ by $(\Qvec-\Ivec)\rvec = \tvec^{\perp},\
\rvec \cdot \evec = 0$, so that
$\Qvec \rvec = \rvec + \tvec^{\perp}$. Now, iterate the
latter to get the identity,
\beq
\Qvec^i \rvec = \rvec  + \sum_{j = 0}^{i-1} \Qvec^j \tvec^{\perp}.
\label{iter}
\eeq
Choosing the arbitrary translation $\yvec_1 = \rvec$
(to put the origin on the axis of the helix)
and eliminating the sum in (\ref{prohelix}) using (\ref{iter})
we have,
\beq
\yvec_{i+1} =  i\, \tvec^{\parallel} +   \Qvec^i \rvec,
\eeq
which, accounting for the conditions $\rvec \cdot \evec = 0$ and
$\Qvec \evec = \evec$, is the equation of a helix.
The orientations $\Rvec_i$ of these helical configurations also
vary in a regular way along the helix in a manner given by
(\ref{helix1}) and illustrated, for example, in Figure
\ref{figure5}.

The basic
geometric information, like formulas for the pitch and radius
of the helix in terms of the given information $(\tvec, \Qvec)$,
can be read off from the formulas given in the preceding
paragraph.
In particular, if $\nu \ge 1$ is the smallest number such that
$\Qvec^{\nu} = \Ivec$, then the pitch is
$\nu | \tvec^{\parallel}|$.
The radius is
$|(\Qvec - \Ivec)^{-1}\tvec^{\perp}|$,
the inverse taken on the plane perpendicular to ${\evec}$.  This
inverse is given by
$(\Qvec - \Ivec)^{-1} =
 \frac{-1}{\text{tr}(\Qvec-\Ivec)} (\Qvec-\Ivec)^T$.

Below, we will observe that the relaxed configurations of
extended or contracted T4 tail can be viewed as a
collection of 6 helices, each with 24 molecules.
To evaluate the positions and orientations of all these
molecules from experimental data, it will be useful to
understand how the orientations of the molecules on a helix
can be varied independently from the shape of the helix.
This is not immediately obvious from the formulas (\ref{helix1}),
but is easy to work out.  First it is clear geometrically
(and it can be shown from the formulas above) that if the
shape of the helix is given, i.e., all of the $\yvec_i$,
then assignment of the orientation of one of the molecules
on the helix determines the orientations of all the others.
Thus there is expected to be one free rotation matrix $\Rvec$
to define this orientation.  Given a helical configuration
defined by $(\yvec_1,\Rvec_1,  \Qvec, \tvec)$, then all
other helical configurations with the same positions are
given by
\beq
   ( \yvec_1, \Rvec_1 \Rvec, \Rvec^T\Qvec\Rvec, \Rvec^T\tvec),
   \ \ \Rvec \in {\rm SO(3)}.  \label{orient1}
\eeq
To use this formula, one can think of beginning with a helix
of the desired shape and then choosing $\Rvec \in {\rm SO(3)}$
so that $\Rvec_1 \Rvec$ is the desired orientation of molecule
1 and all the others then follow.

For sheets there are similar
kinds of helical configurations, having
the shape of a ribbon bent and twisted into a
helix.  These are discussed below in Sections \ref{sect7} and
\ref{sect9}.

\section{Equilibrium, forces and moments}

In this section we work out completely the equilibrium
conditions for chains.

We shall repeatedly have to take a derivative of functions
with respect to a rotation matrix, so we first briefly
explain the meaning of that here.  (It is well-known how
to deal with the fact that rotation matrices lie on the
manifold SO(3) and therefore their components cannot be
varied independently, but we need to explain the notation).
Recall that for an arbitrary skew matrix $\Wvec = -\Wvec^T$
the series $\Ivec + s \Wvec + \frac{1}{2} s^2 \Wvec^2 + \dots$
(i.e., the quantity $s \Wvec$ substituted formally into the
exponential series) converges for all $s$ and yields a
rotation matrix.  For a smooth function $f(\Rvec)$ defined on SO(3) we
extend $f$ smoothly outside SO(3) in any way, and consider
\beq
\frac{d}{ds} f((\Ivec + s \Wvec + \frac{1}{2} s^2 \Wvec^2 + \dots)
\Rvec)\Big{|}_{s = 0}  = \frac{\partial f (\Rvec)}{\partial \Rvec}
\cdot \Wvec \Rvec =  \frac{\partial f (\Rvec)}{\partial \Rvec} \Rvec^T
\cdot \Wvec.   \label{rots1}
\eeq
In 3D $\Wvec = -\Wvec^T$ has an axial vector $\Wvec \avec =
\wvec \times \avec, \forall \avec \in \RR^3$, so the latter
is a linear function of $\wvec$.  We use the notation
$\partial f / \partial \wvec$ for this linear function, i.e.,
\beq
\frac{\partial f (\Rvec)}{\partial \Rvec}\Rvec^T \cdot \Wvec
 =  \frac{\partial f (\Rvec)}{\partial \Rvec}\Rvec^T \cdot (\wvec \times)
\ \stackrel{\rm def}{=}\ \frac{\partial f(\Rvec)}{\partial \wvec}
 \cdot \wvec.     \label{rots2}
\eeq
In rectangular Cartesian components,
\beq
  \frac{\partial f(\Rvec)}{\partial w_j}  = \eps_{ijk}
  \frac{\partial f(\Rvec)}{\partial R_{im}} R_{km}\ ,
   \label{rots3}
\eeq
where $\eps_{ijk}$ is the permutation symbol.

We suppose that
the chain is loaded at its ends by generalized forces
$\fvec_1, \fvec_N$ conjugate to $\yvec_1, \yvec_N$
and $\Mvec_1, \Mvec_N$ conjugate to $\Rvec_1, \Rvec_N$
applied, respectively,  to molecules $1$ and $N$.
The total free energy of the chain and loading devices is then,
\beqs
\lefteqn{\Psi(\y_1,\Rvec_1\, , \ldots,\, \y_N, \R_N) - \yvec_1 \cdot \fvec_1
- \yvec_N \cdot \fvec_N - \Rvec_1 \cdot \Mvec_1 -
\Rvec_N \cdot \Mvec_N}  \nonumber \\
& & = \sum_{i=1}^{N-1} \vphi(\tvec_{i},\Qvec_{i})
 -\ \yvec_1 \cdot \fvec_1
- \yvec_N \cdot \fvec_N - \Rvec_1 \cdot \Mvec_1 -
\Rvec_N \cdot \Mvec_N.
\label{tfe}
\eeqs

According to general principles, the first variation of the
total free energy with respect to rigid translations is the
{\it balance of forces} and its first variation with respect
to rigid rotations is the {\it balance of moments}.  The former,
i.e., the derivative of (\ref{tfe}) with respect to $s$ at
$s = 0$ of a variation $\yvec_i(s)  = \yvec_i + s \cvec$,
for arbitrary choices of $\cvec$, yields, by the frame-indifference
of the quantities $\tvec, \Qvec$,
\beq
\fvec_1 + \fvec_N = 0, \label{balfor}
\eeq
indicating, as expected, that $\fvec_1$ and $\fvec_N$ are
interpretable as simple applied forces.  Similarly, the
first variation $\yvec_i(s)  = (\Ivec + s \Wvec  + \dots) \yvec_i,\
\Rvec_i(s)  = (\Ivec + s \Wvec  + \dots) \Rvec_i$
with $\Wvec = -\Wvec^T$ independent of $i$ yields, using the
arbitrariness of $\Wvec$ and the formulas
(\ref{rots1})-(\ref{rots3}),
\beq
(\yvec_1 - \yvec_N)\times \fvec_N + \mvec_1 + \mvec_N = 0,
\label{balmom}
\eeq
where (in the notation of (\ref{rots2}))
\beq
 \mvec_1 = -\frac{\partial (\Mvec_1 \cdot \Rvec)}{\partial \wvec},\ \
 \mvec_N = -\frac{\partial (\Mvec_N \cdot \Rvec)}{\partial \wvec},
 \label{Mm1}
\eeq
evaluated, respectively at $\Rvec = \Rvec_1, \, \Rvec_N$.
In rectangular Cartesian components,
\beq
 m_{1\, j} = -\eps_{ijk}  M_{1\, im}\,  R_{1\, km}, \ \
  m_{N\, j} = -\eps_{ijk}  M_{N\, im}\,  R_{N\, km}.   \label{Mm2}
\eeq
Clearly, (\ref{balmom}) is a balance of moments
and $\mvec_1, \mvec_N$ are pure moments applied to
molecules $1$ and $N$, these being related by the somewhat
nonobvious formulas to the generalized ``forces''
 $\Mvec_1, \Mvec_N$.
 
Having established these interpretations we now take a general
first variation of the total free energy,
 $\yvec_i(s)  =  \yvec_i + s \uvec_i,\
\Rvec_i(s)  = (\Ivec + s \Wvec_i  + \dots) \Rvec_i$,
where now the positions and orientations of molecules
are varied independently.  The previous variations being
perfectly legitimate, we may as well assume that the
generalized forces satisfy (\ref{balfor}) and
(\ref{balmom}), in which case we write the total free
energy as,
\beq
 \sum_{i=1}^{N-1}
\vphi\left(\Rvec_i^T (\yvec_{i+1} - \yvec_i),
\Rvec_i^T \Rvec_{i+1}\right)
 +\ (\yvec_N - \yvec_1) \cdot \fvec_1
- \Rvec_1 \cdot \Mvec_1 -
\Rvec_N \cdot \Mvec_N.
\label{tfe2}
\eeq
Inserting these variations, differentiating with respect
to $s$, evaluating at $s = 0$ and using the arbitrariness
of $\uvec_i, \Wvec_i$ we get after some manipulation, for interior
molecules,
\beqs
\Rvec_i \frac{\partial \vphi(\tvec_i,\Qvec_i)}{\partial \tvec} &=&
\fvec_N \, , \nonumber \\
(\yvec_{i+1} - \yvec_{i}) \times \fvec_N - \mvec_{i-1,i} +
\mvec_{i,i+1} &=& 0 , \ \ i = 2, \dots, N-1,
\label{intequil}
\eeqs
where $\mvec_{\ell,\ell+1}$ is given compactly by using
the notation (\ref{rots2})-(\ref{rots3}):
\beq
 \mvec_{\ell,\ell+1} = \Rvec_{\ell}
 \frac{\partial
 \vphi \left(\tvec_{\ell}, \Qvec_{\ell}\right)}{\partial \wvec}.
 \label{momel}
\eeq
The argument leading to   (\ref{intequil})$_2$,
(\ref{momel}) is somewhat involved, so we present the
details in the Appendix.

For boundary molecules we have
\beqs
(\yvec_{2} - \yvec_{N}) \times \fvec_N  +
\mvec_{1,2} + \mvec_N  &=& 0 , \ \  \nonumber \\
\mvec_{N-1,N} + \mvec_{N} &=& 0.  \label{bdryequil}
\eeqs
From these equations we interpret  $\mvec_{\ell,\ell+1}$
as the moment on molecule $\ell+1$
produced by molecule $\ell$, and $\Rvec_{\ell}\, {\partial
\vphi(\tvec_{\ell},\Qvec_{\ell})}/{\partial \tvec}$ as the force on
molecule $\ell $ exerted by molecule $\ell +1$, and, even though
the latter is the force on molecule $\ell$,  we should interpret
this force as acting at the center of mass of molecule $\ell+1$.

Helical configurations have a special status with regard to the
interior equilibrium equations.  Recall that these configurations
are defined
by saying that the relative translation and orientation are independent
of the molecule, and they are characterized by (\ref{helix0})
and (\ref{helix1}).  Let $(\tvec, \Qvec)$ be the given relative
translation and orientation associated with a helical configuration.
It follows from the first of (\ref{intequil}) that $\Rvec_i^T \fvec_N$
is independent of $i$ and then, using the first of (\ref{helix1}),
that $\Rvec_1^T \fvec_N$ is on the axis of $\Qvec$:
\beq
 \Qvec \cvec = \cvec, \ \ {\rm where}\ \cvec = \Rvec_1^T \fvec_N.
 \label{helaxis}
\eeq
From the results of Section \ref{sec5}
$\Rvec_1 \cvec$ is the axis of the helical chain.  Thus, (\ref{helaxis})
 has the interpretation that {\it helical configurations are consistent
with axial forces only}.  One can think of this physically in the following
way: given an applied  force $\fvec_N$, the overall rotation $\Rvec_1$
will adjust itself to make the force axial.
It then follows that the $N-3$ interior
equilibrium equations (\ref{intequil})$_1$ collapse to the single
vector equation
\beq
\frac{\partial \vphi(\tvec,\Qvec)}{\partial \tvec}  =
\cvec = \Rvec_1^T \fvec_N.  \label{helforce}
\eeq
Now we turn to the second equilibrium equation (\ref{intequil})$_2$ for
the moments.  Using  (\ref{momel})  and
(\ref{helaxis}), we get that the $N-3$ interior equations
(\ref{intequil})$_2$ collapse to the single
vector equation
\beq
\tvec \times \cvec + (\Ivec - \Qvec^T)\frac{\partial \vphi (\tvec, \Qvec)}
{\partial \wvec} = 0.   \label{helmom}
\eeq
Note that the $\cvec$-component of the latter is an identity, so
(\ref{helmom}) comprises effectively two equations
which relate the applied force to the projection of the moment
on the plane perpendicular to $\cvec$.  The $\Rvec_1\cvec$-component
of (\ref{intequil})$_2$ requires
$(- \Rvec_1^T\mvec_{i-1,i} + \Rvec_1^T\mvec_{i,i+1})\cdot \cvec = 0 ,
\ \ i = 1, \dots, N-1$.  Therefore from the
boundary equation (\ref{bdryequil})$_1$ the
$\cvec$-component of the moments is defined by,
\beq
\left( \frac{\vphi (\tvec, \Qvec)}{\partial \wvec} + \Rvec_1^T \mvec_N \right) \cdot \cvec =0.
\label{axialmom}\eeq
 
Mathematically, we can view the  equation
(\ref{helaxis})
in the following way.  Given the applied force $\fvec_N$,
the magnitude of $\cvec$ is determined by $\cvec = \Rvec_1^T \fvec_N$.
Then, the unknown $\Qvec$ determines the direction of $\cvec$
as well as $\Rvec_1$, up to an arbitrary axial rotation. In addition,
we have freedom to prescribe the applied axial moment $\R_1^T\mvec_N\cdot\cvec$.
We then recognize that the equilibrium equations
(\ref{helforce}), (\ref{helmom}) and (\ref{axialmom})
comprise 6 equations for the determination of the 6 unknowns
$(\tvec, \Qvec)$.
Depending on the convexity properties of $\vphi$,
these are expected to determine $(\tvec, \Qvec)$,
but we do not explore this here.

\section{Bacteriophage T4 tail sheath} \label{sect7}

In this section we specialize the formulas given above
to T4 tail sheath.  The first task is to describe the
sheath in its extended and contracted configurations
and identify the positions and orientations using experimental
measurements.

T4 tail sheath can be viewed as a protein sheet as defined
above.  We can think of a cylinder oriented vertically.
The lowest annulus of the cylinder is a circle of 6 molecules.
Each of these 6 molecules generates a right handed
helical chain consisting of 23 molecules \footnote{Leiman et al.  demonstrate \cite{leiman}
that the tail sheath consists of 23 annuli, rather than
24 as described in preceding papers.  According to this work
the missing annulus can now be ascribed to the baseplate,
based on its detailed protein structure.}.
Hence, we will identify the molecules
accordingly, $(\yvec_{i,j},\Rvec_{i,j}),
\ \ i = 1, \dots, 6,\  j = 1, \dots, 23$.

We work in the usual orthonormal basis $(\evec_1, \evec_2, \evec_3)$
and without loss of generality we will choose the overall
rotation and translation so that the axes of all the helices
coincide with $\evec_3$ and the first annulus lies in the
$\evec_1,\evec_2$ plane.  The first annulus is a circle
(i.e., a degenerate helical chain).
According the results in Section
\ref{sec5}, this case corresponds to the case where $\tvec = \tvec_0$
is perpendicular to the axis of $\Qvec$ and a little calculation
shows that without loss of generality (by suitably rotating the
six molecules about the $\evec_3$ axis) we can assume,
\beq
  \tvec_0 = (-\rho,0,0),\ \ \ \Qvec_{\theta} =
\left( \begin{array}{ccc}
\cos \theta & -\sin \theta & 0 \\ \sin \theta & \cos \theta & 0 \\
 0 & 0 & 1 \end{array} \right),
\label{annulus1}
\eeq
$\rho>0$, and the fact that the 6 molecules are equally spaced on the
helix gives $\Qvec_{\theta}^6 = \Ivec \Longrightarrow \theta = \pi/3$.
Without loss of generality we write
 $\yvec_1 = \rho (1/2, \sqrt{3}/2,0)$.
The radius of the circle of positions is $\rho$.

Emanating from each of these six molecules is a helical chain
whose first molecule has now a given position and orientation.
According to results of Section \ref{sec5}, we need to specify
$(\tvec, \Qvec)$ for each of these chains.  In fact all of these chains
have the same  $(\tvec, \Qvec)$ because suitable rigid rotations
and translations bring them into coincidence with each other:
the whole configuration
of the tail sheath has 6-fold
symmetry.  The axis of $\Qvec$ is again $\evec_3$ so $\Qvec$ has
the form (\ref{annulus1})$_2$, $\Qvec = \Qvec_{\gamma}$.  Thus,
besides the radius $\rho$ of the cylinder,
we need to determine the four parameters
\beq
  \gamma \ \ {\rm and}\ \    \tvec = (\tau_1, \tau_2, \lambda).
\eeq
For this purpose we first show that $\tau_1, \tau_2$ are determined
by $\rho$ and $\gamma$.  Referring
 to Section \ref{sec5} and using that the initial point is
$\yvec_1 = \rho (1/2, \sqrt{3}/2,0)$,  we have from
the equations $\tvec = \tvec^{\parallel} + \tvec^{\perp},
(\Qvec-\Ivec) \rvec = \tvec^{\perp}, \rvec = \yvec_1$
that $(\Qvec-\Ivec)\yvec_1  = \tvec^{\perp}$,
from which $\tau_1, \tau_2$ are given by
\beq
 \tau_1 = \frac{\rho}{2}\left((\cos \gamma-1) -
 \sqrt{3}\sin \gamma  \right), \ \ \
  \tau_2 = \frac{\rho}{2}\left(\sin \gamma +
   \sqrt{3}(\cos \gamma-1)  \right)
\eeq
in terms of $\rho$ and $\gamma$.
It remains to determine $\rho, \lambda,$ and $\gamma$.  The values
of these depend on whether we consider extended or contracted
sheath.

\subsection{Extended tail sheath}  \label{sect7.1}

\begin{figure}
\includegraphics[width=.45\columnwidth]{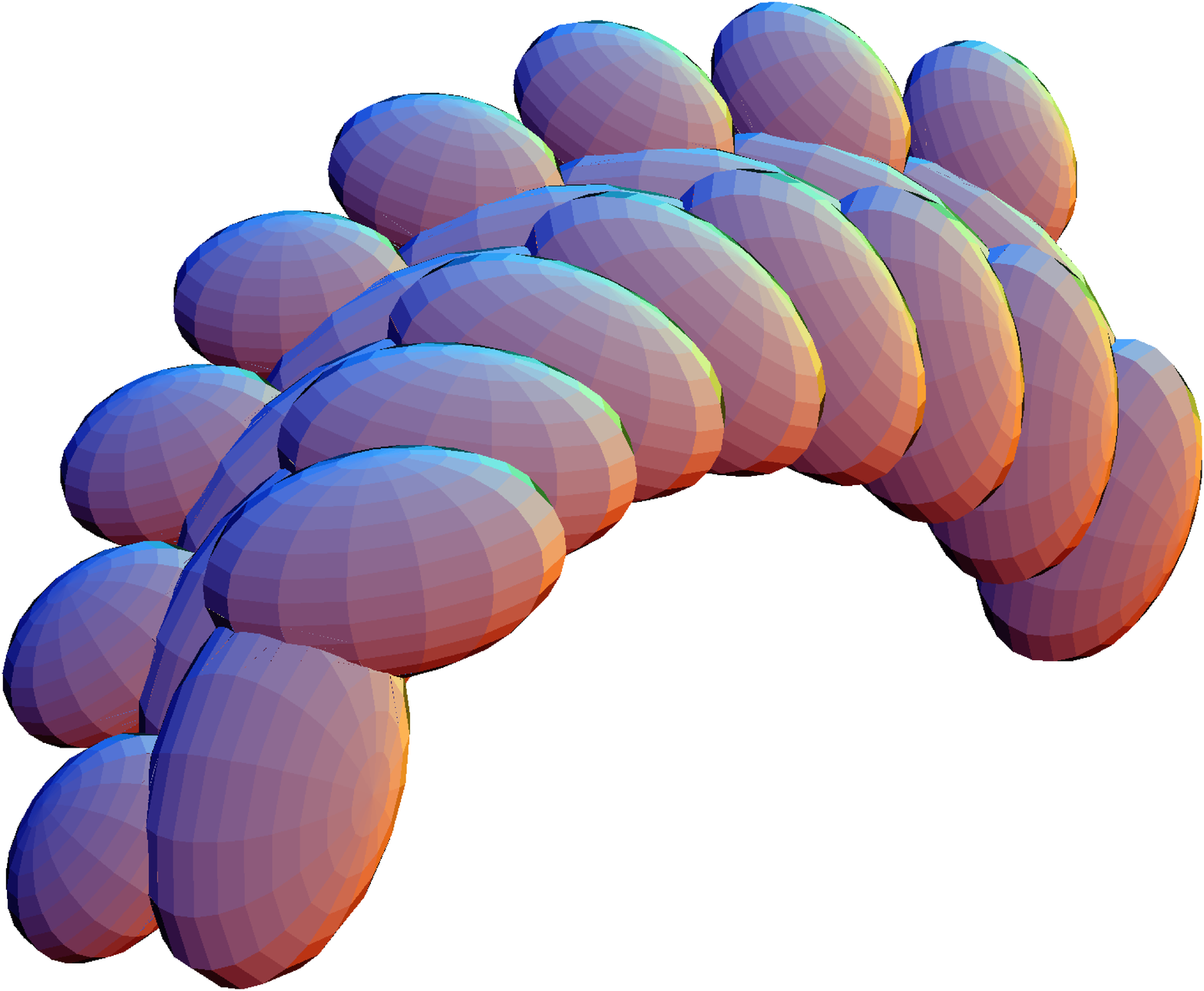} 
\hfill 
\includegraphics[width=.45\columnwidth]{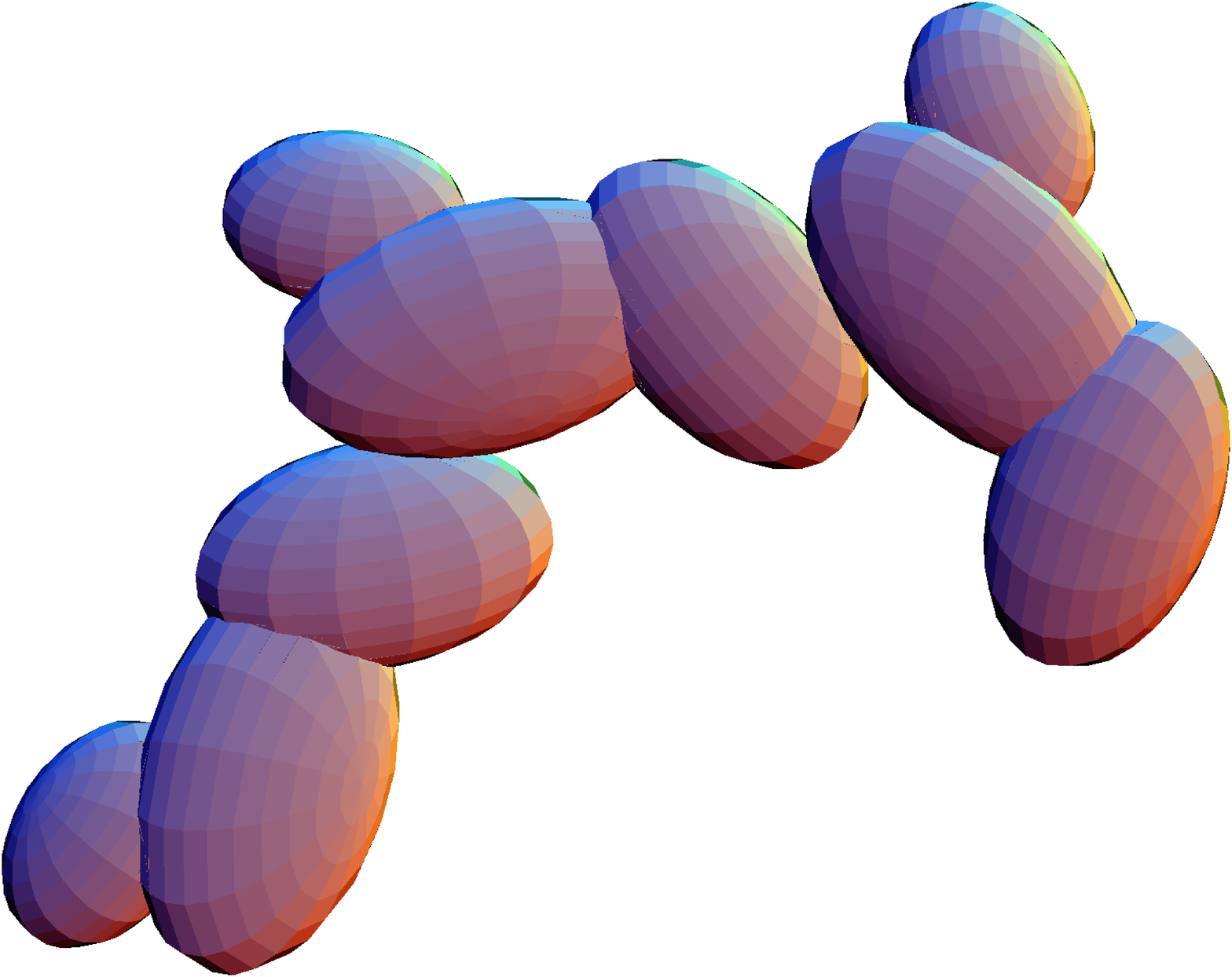}
\caption{Illustration of the 8/3 rule of extended sheath.  Left: the first 8
molecules ($i = 1,\  j = 1, \dots, 8$) on the main helix,
viewed down the axis of
the cylinder.  Right: the first 3 molecules
 ($i = 1, 2, 3,\  j = 1$) on the first
annulus, again viewed down the axis. The slight touching of domains
of neighboring molecules on the right picture is an artifact of
the ellipsoidal approximation; in reality these do not touch.}
\label{figure3}
\end{figure}

Extended sheath has an interesting geometric property
that we term the 8/3 rule.  The rule is that
the 8$^{\rm th}$ molecule along one of these helices, beginning
at a molecule on the first annulus, lies directly over the
third molecule away counterclockwise along the annulus,
Figure \ref{figure3}. (The justification of this rule from
measured data of Leiman et al. ~\cite{leiman} is
given at the end of this section.)
Specifically, in
our notation,
\beq
   \yvec_{1,8} \cdot \evec_1 =  \yvec_{3,1} \cdot \evec_1, \ \ \
   \yvec_{1,8} \cdot \evec_2 =  \yvec_{3,1} \cdot \evec_2,
   \label{8/3rule}
\eeq

This statement and the helical structure of the tail imply
full periodicity,
$\yvec_{(i,j+7)} \cdot \evec_{1,2} =
 \yvec_{(i+2,j)} \cdot \evec_{1,2}$ whenever these are defined.
With regard to the present theory, all the good properties of
helical configurations discussed above (and below)
would hold without this
``accidental'' periodicity.  This suggests that its presence is
perhaps related to something other than the function of the tail,
possibly its self-assembly via annulus-by-annulus epitaxial
growth.  In this regard, if one omits the last
annulus, then the rest of the tail is exactly 1 period.
In other words, without omissions, the 22$^{\rm rd}$ annulus
lies directly over the 1$^{\rm st}$ annulus.   It will be seen
below that this 8/3 rule also applies to the orientation,
$\Rvec_{(1,8)} = \Rvec_{(3,1)}$.   This is of course the
smallest period exhibited by the tail sheath.
Possibly these facts are related to process by which the tail tube directs
the growth of the tail sheath, which is
assembled in the extended form (Below, the contracted form
will not have this or a shorter period).

The two equations (\ref{8/3rule}) give apparently two restrictions on the
remaining  parameters $(\rho, \lambda, \gamma)$.  Written
out using (\ref{helix1}), these two conditions are
\beq
 -\sum_{j = 0}^{6} \Qvec_{\gamma}^j\tvec + \tvec_0 + \Qvec_{\pi/3} \tvec_0
 \parallel \evec_3.
\eeq
In fact this condition only
involves $\gamma$ and is equivalent
to the pair of equations
\beq
2 + \cos 7 \gamma - \sqrt{3} \sin 7\gamma = 0, \ \ \
\sqrt{3} \cos 7\gamma +  \sin 7\gamma = 0.
\eeq
These equations have simultaneous roots at $\gamma = 2 \pi/21 +
2 \pi\, n/7$
where $n$ is an integer.  The root of interest (i.e., corresponding
to a fraction of a turn in the counter clockwise sense)
is $\gamma = 2 \pi/21 $.  From the form of $\tvec$ it can now
be seen that $21\, \lambda$ is the pitch of the helices.

\begin{figure}
\includegraphics[width=.9\columnwidth]{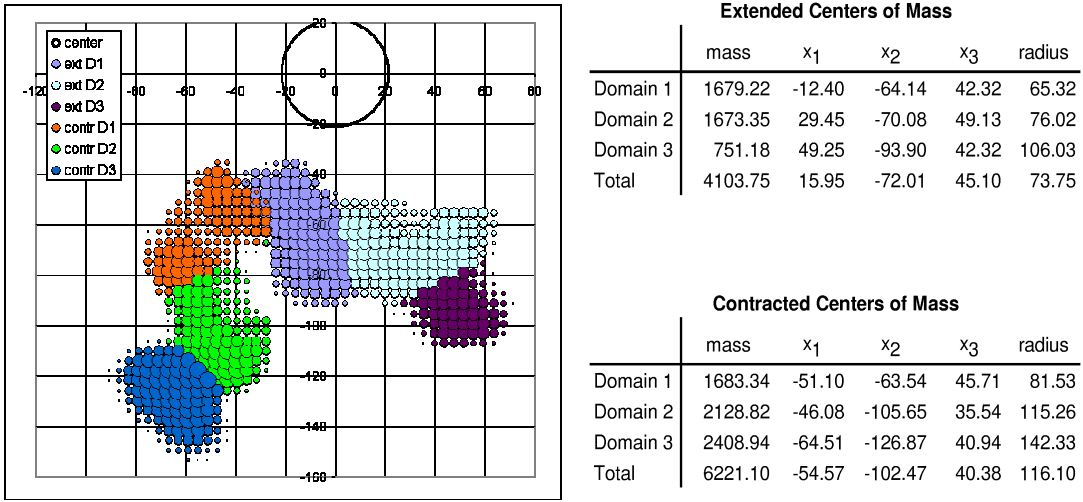}
\caption{Domain coordinates used to determine orientation, $\Rvec_{1,1}$.
The circle denotes the centerline of the cylindrical sheath
and each molecule is modeled by three domains (The three
domains to the right represents a molecule of extended sheath).  Total
mass and center of mass of each domain are shown to the right.}
\label{figure3.5}
\end{figure}

It remains to prescribe the orientations of all the molecules.
As explained in the few lines preceding
(\ref{orient1}) this is assignable independently of the positions.
Since we have put $\Rvec_1 = \Ivec$, we may give this by giving
the orientation of molecule $(1,1)$, from which all the orientations
of all the molecules are determined.  In summary, the following
information is needed from experiment for extended and contracted
tail sheath:
\begin{itemize}
\item The orientation of molecule (1,1) $ = \Rvec_{1,1} \,$;
\item The radius of the cylinder of centers of masses $ = \rho \,$;
\item The pitch of the helices $ = 21\, \lambda \,$.
\end{itemize}
We obtained these from electron density maps of
Leiman et al. \cite{leiman1} (We are grateful to Petr Leiman
for the prepublication data on extended sheath, without
which the present theory would be incomplete).  See
Appendix B for how this data was used to represent the
molecules.  Briefly,
this data does not give atom positions, but gives an
excellent picture of relatively rigid collections of atoms
called domains.  Both extended and contracted sheath consist
of three such domains.  We assumed charge neutrality and
computed centers of mass of domains, then used the formulas
(\ref{ypos}),
(\ref{defgrad}), and (\ref{pd}) to compute the position
and orientation.
Three issues should be noted:
1) with three domains $\Fvec$  is singular with
rank equal to two; nevertheless, $\Rvec$ is uniquely
determined by
(\ref{pd}). 2) This data was rotated about the axis of the
helix and
translated into the position of molecule (1,1) ).
This gives, $\rho = 73.75$ \AA ,  $\lambda = 40.6$ \AA.
We chose extended sheath to be the reference
configuration so that $\Rvec_{1,1} = \Ivec$.
For contracted sheath  $\Rvec_{1,1}$ is given by
(\ref{r11}) below.  3) Note from Figure~\ref{figure3.5} that
masses of domains are not conserved.  This is a consequence of
the flexibility of certain bonds, which causes some
mass to be lost by the averaging procedure inherent in any 3D 
reconstruction. To give definite results we ignored this
problem and used
the measured masses of each domain.
 
\begin{figure}
\includegraphics[width=.35\columnwidth]{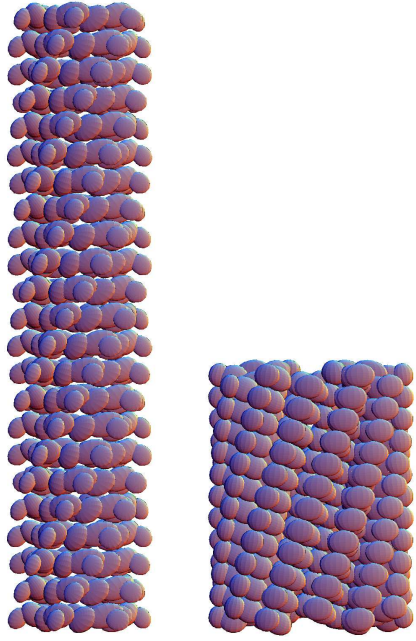}
\caption{Pictures of extended and contracted tail sheath
based on the formula (\ref{T4sheath}), using the method
of visualization described in Section \ref{sect7.2}.}
\label{figure4}
\end{figure}

\subsection{Contracted tail sheath}    \label{sect7.2}

For contracted tail sheath the evaluation is completely analogous
to the above except that the 8/3 rule is replaced by a 12/1 rule,
\beq
   \yvec_{1,12} \cdot \evec_1 =  \yvec_{1,1} \cdot \evec_1, \ \ \
   \yvec_{1,12} \cdot \evec_2 =  \yvec_{1,1} \cdot \evec_2.
   \label{12/7rule}
\eeq

As above this leads to a pair of equations for $\gamma$,
\beq
1- \cos(11 \gamma)  +
        \sqrt{3} \sin(11 \gamma)  = 0, \ \
\sqrt{3} -  \sqrt{3}\cos(11 \gamma)
       - \sin(11 \gamma)  =  0,
\eeq
having simultaneous first positive root at $\gamma = 2 \pi/ 11$.
As above, to complete the description, we need the orientation
of the first molecule $\Rvec_{1,1}$,
the radius of the cylinder $\rho$, and the pitch of the
helices, which in this case is $11 \lambda$.  Using the
electron density maps of Leiman \cite{leiman} in the same
manner as above, we get for contracted sheath
$\rho = 116.1$ \AA ,  $\lambda = 16.4$ \AA, and
\beq
   \Rvec_{1,1} = \left( \begin{array}{ccc}
0.426   & 0.4388 & -0.791  \\
-0.4378 & 0.8653 & 0.244 \\
0.7916 & 0.242 & 0.561
 \end{array} \right)     \label{r11}
\eeq

In summary, the configuration of extended or contracted tail sheath is given
by the following equations:
\beqs
\Rvec_{i,j} &=& \Qvec_{\pi/3}^{i-1} \Qvec_{\gamma}^{j-1}
\Rvec_{1,1}, \nonumber \\
\yvec_{i,j} &=&  \yvec_1  + \sum_{k = 0}^{i-2}\Qvec_{\pi/3}^k\tvec_0
+ \Qvec_{\pi/3}^{i-1}\sum_{k=0}^{j-2}\Qvec_{\gamma}^k \tvec,
\nonumber \\
& & i = 1, \dots, 6,  \ \ \ j = 1, \dots, 23,
\label{T4sheath}
\eeqs
where $\gamma = 2 \pi/21$ for extended and
$\gamma = 2 \pi/11$ for contracted tail sheath.  Here,
$\Qvec_{\theta}, \tvec_0$ are defined by (\ref{annulus1}),
$\yvec_1  = \rho (1/2, \sqrt{3}/2,0)$,
 $\tvec_0 = (-\rho,0,0)$,
$\tvec = \lambda \evec_3 + (\Qvec_{\gamma}-\Ivec)\yvec_1$
and we use the convention
$\Qvec_{\gamma}^{0} = \Ivec$
 (Also, sums of the form $\sum_{k = 0}^m$ where $m<0$ are simply
 put equal to zero).
Pictures of extended and contracted tail sheath obtained
from formulas (\ref{T4sheath}) with the data given above
are shown in
Figure \ref{figure4}.
The method of visualization is to approximate the
domains of the molecules of extended and contracted sheath by ellipsoids, centered
at the centers of mass of the domains, as described in
detail in the Appendix B.  As can be seen there, this is quite
an accurate representation of the molecule.
Then we applied the formulas
(\ref{T4sheath}) to this collection \footnote{Note that the formulas
(\ref{T4sheath}) can be applied with accuracy to any model of the
molecule, including the all atom distribution.  However, it is
worth noting that the striking simplicity of (\ref{T4sheath})
follows the precise definitions of position and orientation
we have chosen; other kinematic descriptions may not give
such simple formulae.}.

\begin{figure}
\includegraphics[width=.35\columnwidth]{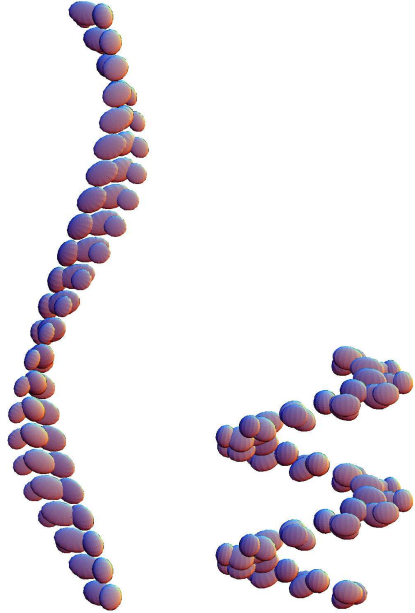}
\caption{The main helix of T4 tail sheath in extended
and contracted forms, illustrating the screw action.}
 \label{figure5}
\end{figure}

There is a substantial screw action that occurs when the
sheath fully contracts.  This can be seen from Figure \ref{figure5}
which shows
the corresponding main right handed helices in extended
and contracted sheath.  If the baseplate is held fixed during
contraction, the neck experiences almost a full turn, the
angle change being about $343$ degrees.

We should add that the data of Leiman et al. ~\cite{leiman} also provides
a direct measure of the validity of the 8/3 and 12/1 rules,
which we have used above to evaluate $\gamma = 2\pi/21 =
17.14^{\circ}\ {\rm and}\ \gamma = 2\pi/11 = 32.73^{\circ}$, respectively.
The direct measurement of Leiman et al. gives the very nearby values
$\gamma = 17.2^{\circ}, 32.9^{\circ}$.
 
\section{A simple constrained theory for
bacteriophage T4 tail sheath}  \label{sect8}

\subsection{Constraints}

Our general expression for the free energy (\ref{Psisheet})
of a protein sheet can be quite complicated, and in our
case is made more complicated by the presence of additional
bonding directions, as we explain below.  In this section
we use the known configurations of the sheath to make
simplifying assumptions that allow us to arrive at a manageable
form of the energy.
\begin{figure}
\includegraphics[width=.9\columnwidth]{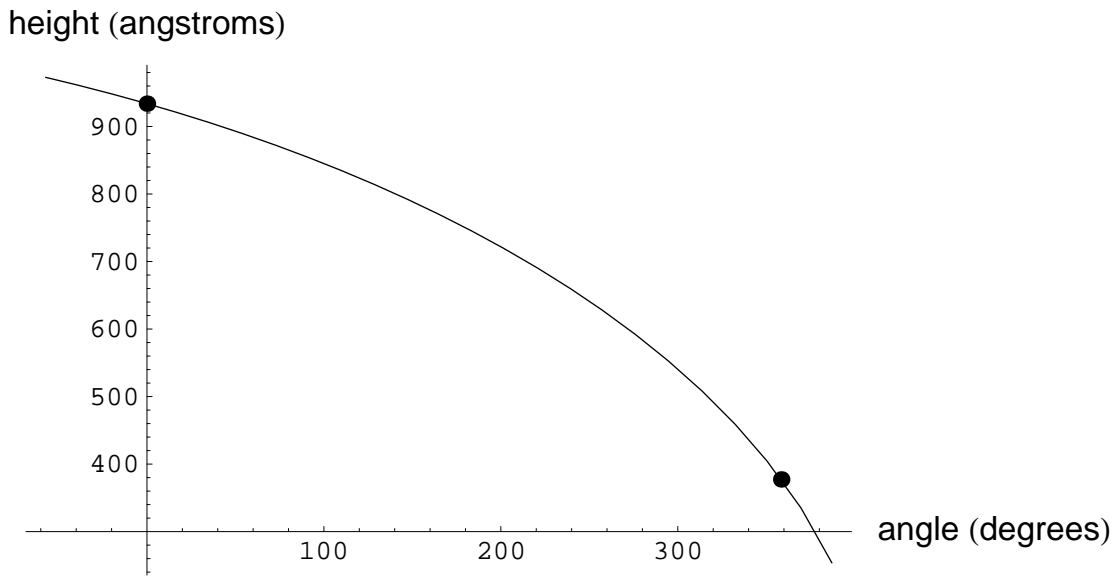}
\caption{Height of the tail sheath vs. end angle (measured from
extended sheath) according
to the constraint, showing a strong first-order Poynting
effect.  Dots
correspond to extended and contracted sheath.}
\label{figure6}
\end{figure}

First we note that each molecule in the sheath undergoes a
substantial motion.  Nevertheless, there are some simplifying
features of this motion.  These features are remarkably close
to the ideas of Pauling \cite{pauling}, who, prior to any knowledge
of T4 tail sheath, theorized that arrays of the helices
of Crane could contract by having adjacent turns of the helix
form bonds.   Later,  in his study of T4 tail sheath,
Moody \cite{M1973} observed that bonds on the right-handed
helix remained to some extent invariant during contraction.
He noticed
that, on the main helix, while there is a substantial
relative rotation of molecules, the distance between neighbors
does not change too much.  This concerns\footnote{The relation
between Moody's notation for bonds and ours is $AB = (i,j)(i,j+1),\
AB' = (i,j)(i-1,j+1),\ AC' = (i,j)(i-1,j+2)$. \label{moodynote}} neighboring
molecules
of the form $(i,j),(i,j+1)$, shown e.g. for $i = 1$ in Figure
\ref{figure5}.  According to our equation (\ref{T4sheath})
we have  for both extended and contracted sheath that this
distance is $|\tvec| = |\Rvec_{i,j}^T (\yvec_{i,j+1}-\yvec_{i,j})|$;
it is of course independent of both $i$ and $j$ and is given by
\beq
|\tvec|^2 =  \lambda^2 + 2\rho^2(1 - \cos \gamma).
\label{fullcon}
\eeq
When this is evaluated for extended and contracted sheath
using the data above
we get, respectively, $|\tvec| = 46.2, 67.4\, $\AA.  While these
are fairly close, Moody noticed that if, instead of using the
separate radii of extended and contracted sheath, one uses
in both cases an effective radius of $\rho = \rho_{\rm eff} =
77.6$\AA, then $\sqrt{\lambda^2 + 2\rho_{\rm eff}^2(1 - \cos \gamma)} =
 46.7\,$\AA\  for both contracted and extended sheath.
The reason for the smaller-than-average effective radius
presumably relates to the relative importance of the bonding
of inner domains, which appear to be in contact in EM
cross-sections of the sheath at a radius near $\rho_{\rm eff}$.
 
We remark that if we approximate $\cos \gamma$ by
$1- (1/2)\gamma^2$ in the expression
$\lambda^2 + 2\rho_{\rm eff}^2(1 - \cos \gamma),$
  and also adjust the value of $\rho_{\rm eff}$
 slightly to $\rho_{\rm eff} = 76.33\,$\AA\ , then we have the following
 simple quadratic condition:
\beq
 \lambda^2 + \gamma^2 \rho_{\rm eff}^2 = \left\{
 \begin{array}{l} 2170\ {\text \AA}^2 \ \  {\rm for\ extended\ sheath,}  \\
 2170\ {\text \AA}^2 \ \ {\rm for\ contracted\ sheath.} \end{array} \right.
 \label{con1}
\eeq
In view of its physical interpretation, we assume (\ref{con1})
represents a special stiffness in T4 tail
sheath and we adopt it as
a constraint for all values of $\lambda$
and $\gamma$.  Below we generalize it to distorted
configurations.

The constraint (\ref{con1}) has an interesting consequence.
To describe this, we first recall that according to
macroscopic nonlinear elasticity, a uniformly twisted cylinder
subject to zero axial force and free sides
changes its diameter and also its length.  The latter is
referred to as the Poynting effect. It is generically
a second order effect: the elongation goes as the square of the
angle of twist of the cylinder; the elongation can
be either positive or negative and it is typically positive
(lengthening) for elastomeric materials.  For uniform states of
T4 tail sheath, that is, states given by the formula
(\ref{T4sheath}) subject to the constraint
(\ref{con1}), we have a very strong {\it first-order} Poynting effect.
That's because, by (\ref{T4sheath}), the the end angle measured
from the extended configuration is $22 \gamma$ while the
height is $22 \lambda$.

The predicted height vs. twist relation
is shown in Figure \ref{figure6}.  This is essentially a plot
of the constraint (\ref{con1}).  We note that if the approximation
$\cos \gamma \approx 1-(1/2) \gamma^2$ is not made, then, on the
scale of Figure \ref{figure6}, the resulting graph is
indistinguishable from Figure \ref{figure6}.
 Note the dramatic Poynting effect,
particularly at contracted sheath.  It would be interesting
to look at this relationship experimentally.

\begin{figure}
\includegraphics[width=.45\columnwidth]{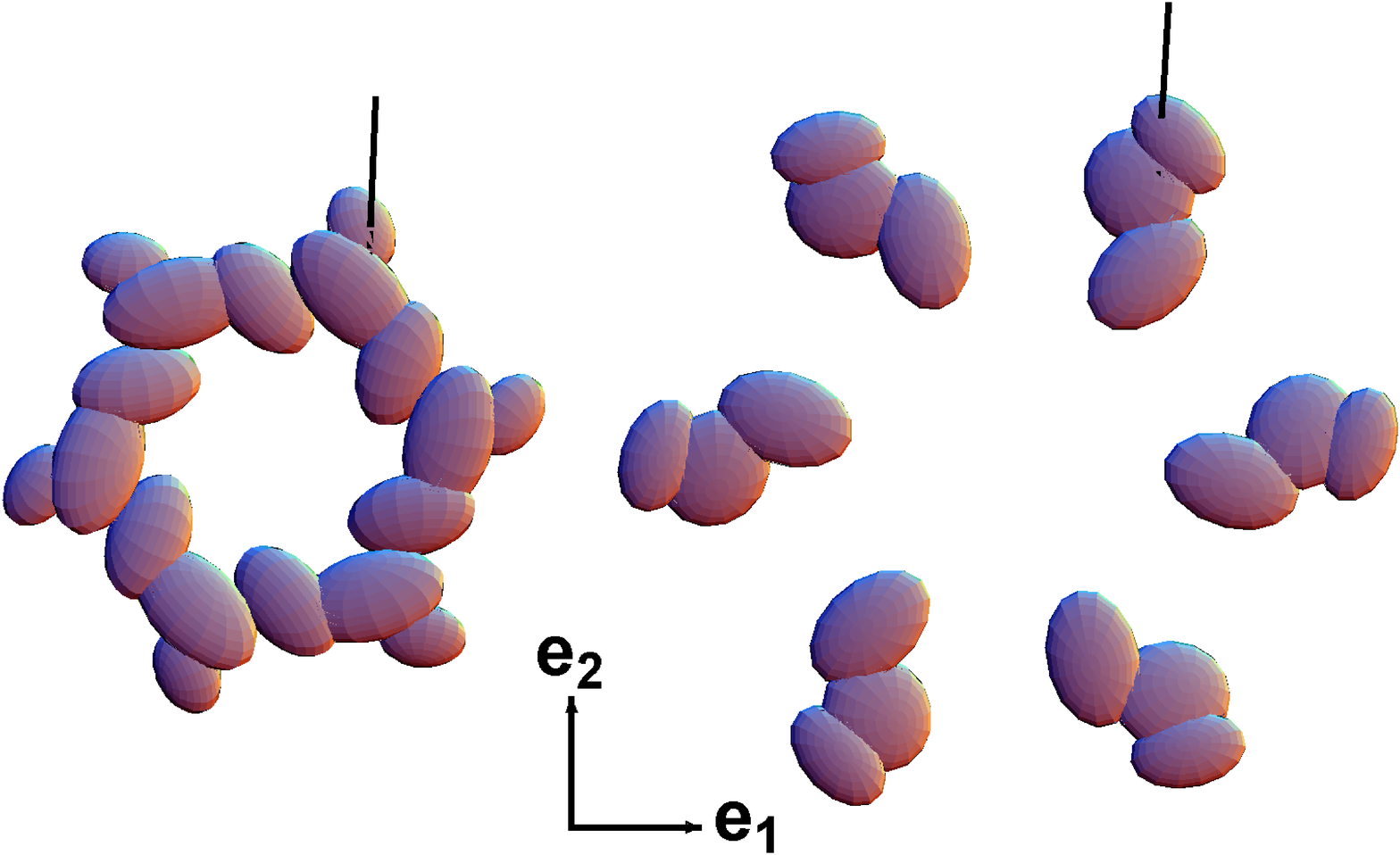}
\hfill
\includegraphics[width=.45\columnwidth]{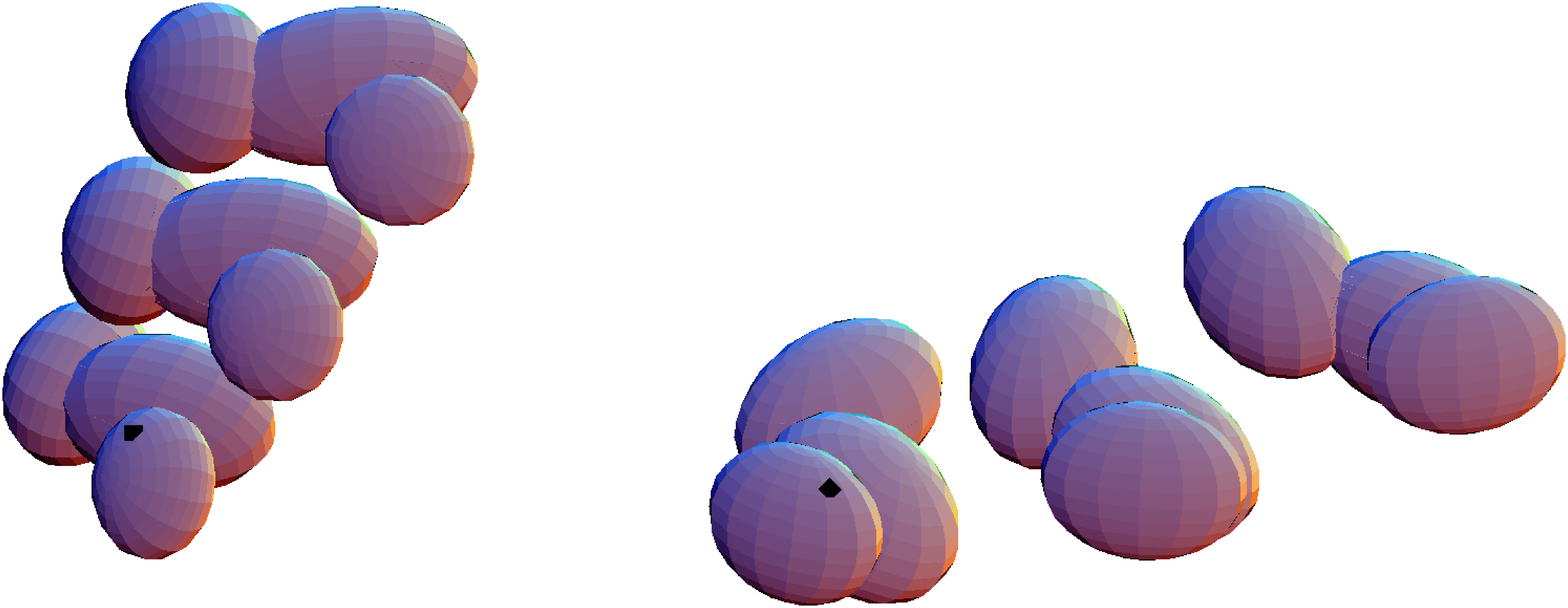}
\caption{Left pair: the first annulus $(i = 1,\dots,6,\ j = 1)$
of extended and contracted sheath viewed down the axis of
the cylinder, with the axis of the rotation
of molecule 1 shown passing through its center of mass.
Right pair: the first three molecules on the main helix
$(i = 1,\ j = 1,2,3)$ of extended and contracted sheath
viewed parallel to the axis of rotation (shown as the
black dot)}
 \label{figure7}
\end{figure}

There is another simplifying feature of the deformation of
T4 tail sheath that concerns the orientation. For molecule
$(1,1)$ the rotation that maps extended to contracted sheath
is given in (\ref{r11}), and its axis is given by
\beq
 (0.001 , 0.875, 0.485).    \label{ax}
\eeq
The angle of rotation is close to
$64.8^{\circ}$.  Remarkably, the axis of rotation
(\ref{ax}) is within about $1^{\circ}$ of
$(0, \sqrt{3}/2, 1/2)$.  A possible reason for this rotation
and its implications become clear when we superimpose the rotation
axis (black line) on pictures of molecule $(1,1)$ of extended
and contracted tail sheath, Figure \ref{figure7}.  From these
pictures, if one thinks of the molecules as having the shape
of a kind of twisted banana, then evidently the axis of
rotation passes through its axis.  Thus, the rotation
of molecules of tail sheath seems largely constrained by
steric hindrance. But there is another feature of this
rotation that is suggested by the two pictures on the
right of Figure \ref{figure7}.  In these two pictures we
are looking directly down the axis of rotation.
One can see that the
rotation of $(1,1)$ of about $60^{\circ}$
is causing it to align itself approximately
with the main helix. As
above, this is consistent with
the idea that there are strong bonds linking molecules
on this helix that not only constrain lengths but also
relative rotations. In fact, even though the molecules
depicted at the right of Figure \ref{figure7} do not touch,
there are strong bonds that link the innermost domains.

We now develop this idea quantitatively. To account for
the evidence for steric hindrance, we assume that
the orientation $\Rvec_{1,1}$ has the fixed axis which
we take to be $(0, \sqrt{3}/2, 1/2)$, but we allow the
angle of rotation to be free for the moment, i.e.,
\beq
   \Rvec_{1,1}(\theta) = \left( \begin{array}{lcc}
\cos \theta  & \frac{1}{2}\sin \theta &
-\frac{\surd{3}}{2}\sin \theta  \\
\, -\frac{1}{2}\sin \theta  & \frac{3}{4} +
\frac{1}{4}\cos\theta &
\frac{\surd{3}}{4} -\frac{\surd{3}}{4}\cos\theta \\
 \frac{\surd{3}}{2}\sin \theta & \frac{\surd{3}}{4}
 -\frac{\surd{3}}{4}\cos\theta  & \frac{1}{4} +
  \frac{3}{4}\cos\theta
 \end{array} \right).  \label{rtheta}
\eeq
Guided by the pictures on the right of
Figure \ref{figure7} and the motivation above, we
compute $\Rvec_{1,1}^T(\theta)(\yvec_{1,2}- \yvec_{1,1})$
for extended and contracted sheath using (\ref{rtheta})
and the corresponding measured values of
$\theta = 0,\ 64.8^{\circ}$ respectively.  The two
vectors obtained are fairly close to each other as is
expected based on Figure \ref{figure7}.  However,
this computation reveals that the projection of
these two vectors on the 1 axis is exceptionally close.
That is, $\evec_1 \cdot  \Rvec_{1,1}^T(\theta)(\yvec_{1,2}-
\yvec_{1,1})$  has nearly the same value of
$-21.2$ \AA\ for extended
and contracted tail sheath.  We again hypothesize that
this represents a special stiffness in this system, and
we adopt it as a constraint,
$\evec_1 \cdot \Rvec_{1,1}^T(\theta)(\yvec_{1,2}- \yvec_{1,1}) = -21.2$
\AA.   Written out using (\ref{rtheta}) and (\ref{T4sheath}),
this constraint is,
\beq
2 \rho \cos \theta \left( 1 + \sqrt{3} \sin \gamma
-\cos \gamma  \right) +
\sin \theta \left( \rho \sin \gamma +  \rho \sqrt{3} \cos \gamma
- \sqrt{3}(2 \lambda + \rho)  \right) = 84.8\ {\text \AA}.
\label{con2}
\eeq
It is natural to use this constraint to solve
for $\theta$, effectively making the orientation of each molecule
slave to the variables that describe the spatial positions
of the sheath.  This is always possible for a wide range of
reasonable values
of $\gamma, \rho, \lambda$ satisfying the earlier
constraint (\ref{con1}).  Some care has to be exercised with
uniqueness, since generically (\ref{con2}) has a pair of
solutions $\theta$; however, only one of these lies
in a modestly expanded
interval containing $[0, 64.8^{\circ}]$.

For uniform states, i.e., configurations obtainable using
the formula (\ref{T4sheath}), the constraints (\ref{con1})
and (\ref{con2}) reduce the energy to a function of the
kinematic variables $\rho$ and $\gamma$, effectively, radius
and twist.  It would be natural now to write the energy as
a double-well energy in $\rho, \gamma$, with wells appropriate
to contracted and extended sheath.  However, it is advantageous
to consider also distorted states, so that process
of transformation can be described.

\subsection{Nonuniform states}

To describe nonuniform states, we first notice that our basic
formula (\ref{T4sheath}) is still useful.  In fact, this
formula can be used to describe an arbitrarily distorted
sheath, by simply allowing $\rho, \gamma, \lambda, \Rvec_{1,1}$
to depend on $(i,j)$. To see this, we  notice that
if the molecule $(i,j)$ occupies a certain position
and has a certain orientation, then one can always find
a helical cylinder with molecule $(i,j)$ in the given
position and with the given orientation.  Effectively, the
formula (\ref{T4sheath}) with
variables $\rho, \gamma, \lambda$ defines certain helical coordinate
system based on the structure of T4 sheath.  We note that
this generalization changes somewhat the geometric interpretations
given above of the variables $\rho, \gamma, \lambda, \Rvec_{1,1}$.

In a setting of this generality, one could make a reasonable
extrapolation of what should be the constraints, based on the
stiffnesses of the main helix discussed above, but the resulting
276 degrees of freedom would still be rather large; once the
energy of T4 sheath becomes known
quantitatively, it will then be worthwhile
doing something like this, since general configurations and forces
could be then computed using standard nonlinear optimization
techniques. For the present, we make a 1-D ansatz that positions
and orientations are the same on each annulus, that is,
\beqs
 (\yvec_{i,j}, \Rvec_{i,j}) \ {\rm is\ given\ by}\ (\ref{T4sheath})
 \ {\rm with}\ \rho = \rho_j,\ \gamma = \gamma_j,\
 \lambda = \lambda_j,\ \Rvec_{1,1} = \Rvec_j, \nonumber \\
 i = 1, \dots,6,\ j = 1, \dots, 23.  \label{1DT4sheath}
\eeqs

Our first goal is to reformulate the constraints in terms of
these variables.  We begin with the first constraint (\ref{con1}).
If we calculate $|\tvec| = |\Rvec_{i,j}^T (\yvec_{i,j+1}-
\yvec_{i,j})|$ using (\ref{1DT4sheath}) we see that it depends
in a somewhat complicated way on $j$, but we can also see from
the expressions that there is a natural change of variables
that restores the simplicity of the expressions for uniform
states.  That change of variables is:
\beq
 \bar{\gamma}_j = j (\gamma_{j+1} - \gamma_j) + \gamma_j,  \ \ \
 \bar{\lambda}_j = j (\lambda_{j+1} - \lambda_j) + \lambda_j
 \ \ \  j = 1, \dots, 22.   \label{newvars}
\eeq
The inverse mapping is simple averaging:
\beq
\gamma_j = \frac{1}{j-1}\left( \bar{\gamma}_1 +
\dots  + \bar{\gamma}_{j-1} \right),\ \
\lambda_j = \frac{1}{j-1}\left( \bar{\lambda}_1 +
\dots  + \bar{\lambda}_{j-1} \right), \ \ j = 2, \dots, 23.
\label{inversenewvars}
\eeq
Note that for uniform states, $\bar{\gamma}_j = \gamma_j = \gamma$
and $\bar{\lambda}_j = \lambda_j = \lambda$.  When the expression
$|\tvec|^2 = |\Rvec_{i,j}^T (\yvec_{i,j+1}-
\yvec_{i,j})|^2$ is evaluated for 1-D states in these new
variables, it becomes,
\beq
 \rho_j^2 - 2 \rho_j \rho_{j+1}\cos \bar{\gamma}_j
 + \rho_{j+1}^2  + \bar{\lambda}_j^2,    \label{1Dfullcon}
\eeq
and the connection with (\ref{fullcon}) is immediately clear.
In fact, it is expected based on the definition of
$\bar{\gamma}_j$ that the approximation
$\cos \bar{\gamma}_j \approx 1- (1/2) \bar{\gamma_j}^2$ is
still reasonable and then (\ref{1Dfullcon}) becomes
$(\rho_{j+1} - \rho_j)^2 + \rho_{j+1} \rho_j \bar{\gamma}_j^2
+\bar{\lambda}_j^2$.   Comparing with (\ref{con1}), it is natural
to again replace all the $\rho_j$ by the effective radius
$\rho_{\rm eff}$.  We therefore adopt in the nonuniform case
the constraint
\beq
 \bar{\lambda}_j^2 + \bar{\gamma}_j^2\, \rho_{\rm eff}^2 =
 2170\ {\text \AA}^2,     \label{1Dcon1}
\eeq
where $\rho_{\rm eff} = 76.33$\ \AA.

Now we generalize the constraints (\ref{rtheta}),
(\ref{con2}) on the orientation.  First, we recall that our
way of writing the formula  (\ref{T4sheath}) automatically
adjusts the orientation of each molecule on the sheath in
a consistent way (preserving the helices) in response to
a change of $\Rvec_{1,1}$.  Since we assumed above
that $\Rvec_{1,1}$
has the same axis for extended and contracted sheath, then
we assume this remains true for nonuniform states
and $\Rvec_{1,1}$ continues to have the form (\ref{rtheta})
with $\theta$ replaced by $\theta_j$.

\begin{figure}
\includegraphics[width=.9\columnwidth]{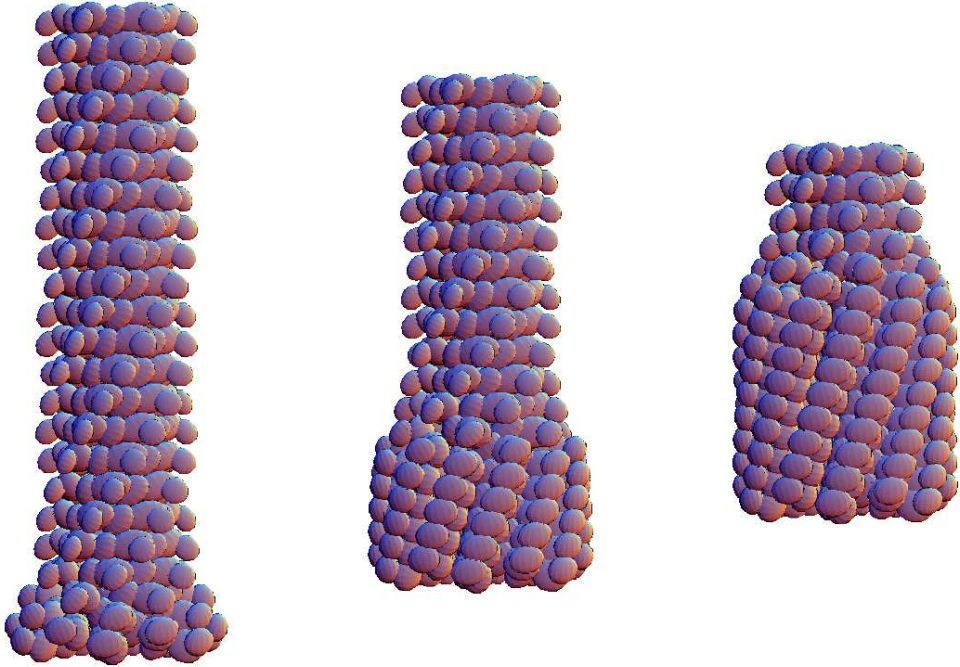}
\caption{Deformations of tail sheath satisfying the constraints
and exhibiting transformation. See text.}
 \label{figure8}
\end{figure}

Once again, the change of variables (\ref{newvars}) proves
to be extremely useful, for if we now calculate the quantity
$\evec_1 \cdot \Rvec_{i,j}^T (\yvec_{i,j+1}-
\yvec_{i,j}) = \evec_1 \cdot \Rvec_{j}^T (\yvec_{i,j+1}-
\yvec_{i,j})$, we get,
\beqs
\lefteqn{ \frac{1}{4}\left( - 2 \cos \theta_j \left( \rho_j +
\rho_{j+1}  \sqrt{3} \sin \bar{\gamma}_j
- \rho_{j+1}\cos \bar{\gamma}_j  \right) \right. }   \nonumber  \\
& & \left. - \sin \theta_j \left( \rho_{j+1} \sin \bar{\gamma}_j
 + \rho_{j+1} \sqrt{3} \cos \bar{\gamma}_j
- \sqrt{3}(2 \bar{\lambda}_j + \rho_j)  \right)
\right),  \label{1Dtj}
\eeqs
with the obvious relation to (\ref{con2}).  We therefore adopt the
following constraint on orientation in the nonuniform case:
\beqs
\lefteqn{  2 \cos \theta_j \left( \rho_j +
\rho_{j+1}  \sqrt{3} \sin \bar{\gamma}_j
- \rho_{j+1}\cos \bar{\gamma}_j  \right)  }  \nonumber \\
& & + \sin \theta_j \left( \rho_{j+1} \sin \bar{\gamma}_j
 + \rho_{j+1} \sqrt{3} \cos \bar{\gamma}_j
- \sqrt{3}(2 \bar{\lambda}_j + \rho_j)  \right)
 = 84.8\ {\text \AA}. \label{1Dcon2}
\eeqs
We again view this as a way to determine
$\theta_j,\ j = 1, \dots, 22$, making the orientation
slave to the other variables.

In summary, there is a natural expression of the constraints
within the context of the 1-D ansatz, this being
(\ref{1Dcon1})-(\ref{1Dcon2}); no internal contradictions
arise, and there is freedom to make a variety of distorted
states that interpolate contracted and extended sheath.
The unconstrained kinematic variables can be taken
to be local twist and radius, which for distorted states
turn out to be  $\bar{\gamma}_1, \dots, \bar{\gamma}_{23}$ and
${\rho}_1, \dots, {\rho}_{23}$.   If these variables are
subject to a simple interpolation between extended and contracted
sheath, by defining
\beqs
 \bar{\gamma}_j &=& \mu(j)\, (2\pi/21) + (1- \mu(j))\,(2 \pi/11), \nonumber \\
{\rho}_j &=& \mu(j)\, 73.75\ {\text \AA} + (1- \mu(j))\,
116.1\ {\text \AA},
\eeqs
where, for example, $\mu(j)$ is a simple ``tanh'' transition layer,
$\mu(s) = \frac{1}{2}(\tanh((s - j_0)/w) +1)$, then one can exhibit a contracting
sheath as is shown in Figure \ref{figure8}.  These pictures are produced in
this way, using the constraints (\ref{1Dcon1}) and (\ref{1Dcon2})
to determine the $\lambda_j$ and $\theta_j$ and then placing
all in the formula (\ref{T4sheath}), as directed by
(\ref{1DT4sheath}).  All three of these pictures have the same
interfacial width $w = 1.5$ and interfacial positions
$j_0 = 4, 12, 20$, respectively.  These are not
necessarily equilibrium
states, as the computation of these would depend on a quantitative
knowledge of the energy function, which we do not yet know.

\begin{figure}
\includegraphics[width=.9\columnwidth]{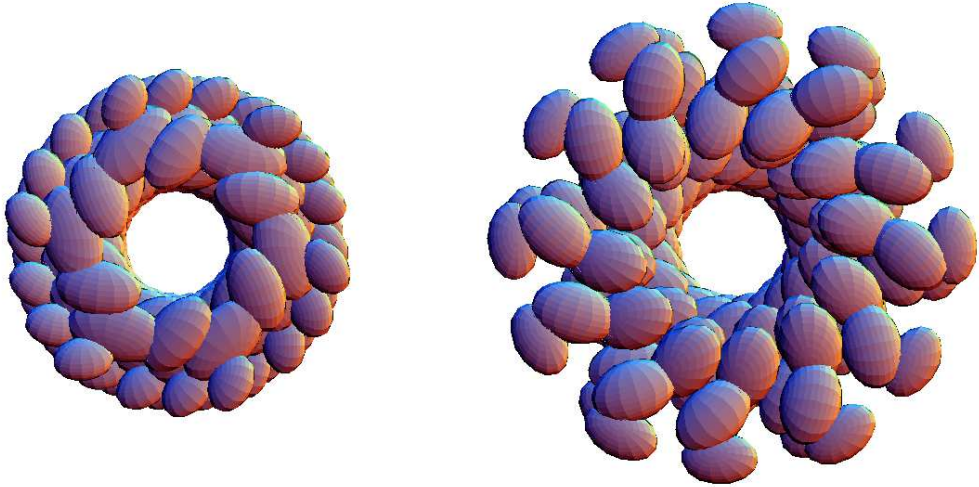}
\caption{Nucleation of the phase transformation in T4 sheath
as viewed from below.  Left: extended sheath. Right: view of
sheath with the first annulus fully transformed, as in the
leftmost picture of Figure \ref{figure8}. }
 \label{figure9}
\end{figure}

These pictures are interesting from the point of view of nucleation.
One of the important issues (raised in \cite{rdj_kfh_00}) is that
T4 tail sheath is at a scale that would seem to suppress martensitic
phase transformation.  Briefly, the argument is the following:
in order to have phase transformation with a distortion one
expects an interface to pass through the body having
a transition layer between phases.  But because of
the scaling between bulk and interfacial energy, the interfacial
energy should dominate at sufficiently small scales, and, therefore,
in a sufficiently small body, one would necessarily
pay more free energy
for the transition layer than the lowering of free energy
due to the presence of the new phase. In the present case
``interfacial'' and``bulk'' energies are better thought of as line
and surface energies, but the argument is
similar.  Thus, nucleation is expected
to be an important issue for phase transitions at small scales,
and this is particularly true in the present case in view
of the enormous transformation strain of T4 tail sheath.
It is known from the work of Moody that transformation begins
at the baseplate.
The distortion of the first annulus upon nucleation can be seen
in Figure {\ref{figure8}.  An alternative
view is seen in Figure \ref{figure9} which shows
a view from below; in this figure the lowest annulus is nearly
fully transformed ($j_0 = 4$) on the left while the corresponding
untransformed sheath is shown on the right.

Finally, a brief remark about constraints and frame-indifference.
It is well-known that internal constraints in mechanical systems
should be frame-indifferent, and this may not be obvious in the
present case. In (\ref{T4sheath}) there is some freedom of how
one assigns a change of frame, either attributing this to changes
of $\gamma, \rho, \lambda, \Rvec_{1,1}$ or, for example,
to changes of $\Qvec_{\gamma}$ (at constant $\gamma$), $\Qvec_{\pi/3}, \tvec_0, \tvec,
\yvec_1, \Rvec_{1,1}$.   The latter is preferred, and also preserves the
1-D ansatz.  The precise form of a change of frame $\yvec \to
\Rvec \yvec + \cvec$, $\Rvec \in$ SO(3)  is then
$\Qvec_{\gamma} \to \Rvec \Qvec_{\gamma} \Rvec^T,
\Qvec_{\pi/3} \to \Rvec \Qvec_{\pi/3} \Rvec^T,
\tvec_0 \to \Rvec \tvec_0, \tvec \to \Rvec\tvec,
\yvec_1 \to \Rvec\yvec_1 + \cvec, \Rvec_{1,1} \to
\Rvec \Rvec_{1,1}$.  With this understanding,
$\gamma, \rho, \lambda$  are objective scalars and the
constraints are frame-indifferent.

\subsection{Free energy}   \label{sect8.3}

Having reduced the complexity of the energy by formulating
constraints for a 1-D ansatz,
we are now in the position to suggest a relatively simple form of the
energy function for nonuniform states of the form
(\ref{1DT4sheath}).   We take the independent variables to
be $\bar{\gamma}_1, \dots, \bar{\gamma}_{23}$ and
${\rho}_1, \dots, {\rho}_{23}$. In the extended T4 tail
sheath the main helix is the only direction of strong bonding;
however, in contracted sheath there are three
bonding directions, as identified by Moody \cite{M1973}
and Leiman et al. \cite{leiman1}.
  These are the bonds $(i,j)-(i,j+1),\
(i,j)-(i-1,j+1),\ (i,j)-(i-1,j+2)$.  For this bonding the
free energy
is a minor generalization of  (\ref{Psisheet}):
\beqs
\lefteqn{\Psi(\y_{1,1},\Rvec_{1,1}\, , \ldots,\, \y_{6, 23}, \R_{6, 23})}
 \nonumber  \\
&=& \sum_{i \in \{1, \dots, 6\},\ j \in \{1, \dots, 23 \} }
\psi_1(\y_{i,j},\R_{i,j},\y_{i,j+1},\R_{i,j+1}) +
\psi_2(\y_{i,j},\R_{i,j} ,\y_{i-1,j+1},\R_{i-1,j+1}) \nonumber \\
& & \quad \quad \quad \quad \quad \quad \
 +\ \psi_3(\y_{i,j},\R_{i,j} ,\y_{i-1,j+2},\R_{i-1,j+2}).
\label{T4energy}
\eeqs
Here we have omitted separate consideration of boundary molecules;
to account for molecules beyond the boundaries, we do a
suitable periodic extension.  Recall that the
$\psi_1, \psi_2, \psi_3$ depend on certain objective quantities,
the $\tvec$'s and $\Qvec$'s,
cf., (\ref{obj_vars}).

We assume the 1-D ansatz (\ref{1DT4sheath}) and the constraints
(\ref{1Dcon1}), (\ref{1Dtj}), (\ref{1Dcon2}). If we
write out all of the frame-indifferent expressions
appearing in the arguments
of $\psi_1$ in the
sum (\ref{T4energy}), we have
\beq \Rvec^T_{i,j} \Rvec_{i,j+1} =
 f( \rho_j, \rho_{j+1},\bar{\gamma}_j) \ \ \ \
\Rvec^T_{i,j} (\y_{i,j+1}- \y_{i,j})
=g( \rho_j, \rho_{j+1},\bar{\gamma}_j),
\eeq
where $f$ and $g$ are somewhat complicated algebraic
vector-valued functions.
We  recall that this bond (along the main helix)
guides the assembly of the extended state and is preserved
throughout contraction.
Since this bond is relaxed in the extended state and undergoes
relatively small deformations,
one simple way to model it is as a harmonic function centered
at the extended state:
\beq
\psi_1(\y_{i,j},\R_{i,j},\y_{i,j+1},\R_{i,j+1})= \frac{1}{12}
\left\{\begin{array}{c}
\rho_j - \rho_e \\ \\
\bar{\lambda}_j - \lambda_e \end{array}\right\} {\bf \cdot}
\left[ \begin{array}{cc}
k_1 & k \\ \\
k & k_2
\end{array}\right]
\left\{\begin{array}{c}
\rho_j - \rho_e \\ \\
\bar{\lambda}_j - \lambda_e \end{array}\right\}
+ k_3 (\rho_{j+1} - \rho_{j})^2 \label{bond1}
\eeq
where $k$ and $k_i>0$ are constants, $\rho_e$ and $\gamma_e$ are the values measured for the
extended sheath.  The term containing $k_3$ is suggested by the
presence of $\rho_j, \rho_{j+1}$ and the expectation that this
energy is minimized by the uniform state: this term is somewhat
like the terms of the energy of a liquid crystal.

The bond $(i,j),(i-1,j+1)$ spans between adjacent main helices.
This bond is largely non-existent in the extended sheath and 
its formation drives the contraction.
However, the energy $\psi_2$ for this bond depends on the
same set of variables
$\rho_j, \rho_{j+1}, \bar{\gamma}_j$ as for $(i,j),(i,j+1)$.
The radius and pitch of adjacent turns of the main helix provides
a measure of the
second bond's state.  If the adjacent turns are
close to the contracted state then
the bond is formed.  For configurations where the helices are
far apart the bond is essentially broken.
And for configurations where the helices become very close
there is a strong repulsion.
Consistent with this we propose the potential
\beq
\psi_2(\y_{i,j},\R_{i,j} ,\y_{i-1,j+1},\R_{i-1,j+1})
= \frac{1}{12}\left( 1-k_4(\rho_j-\tilde{\rho}_c)^2 -
 k_5 (\rho_{j+1}-\rho_j)^2 \right)
{\calL}(\bar{\lambda}_j),
\eeq
where
\beq
\calL(\lambda) = \left\{  \begin{array}{ll}
 -a (c-\lambda)^2(c-3\tilde{\lambda}_c +
2 \lambda), & \lambda \le c, \\
0, & \lambda > c.  \end{array} \right.  \label{calL}
\eeq
is similar in shape to a Lennard-Jones potential, except that it
has a cut-off at $c$, where $\calL$ and its first derivative
vanish (It is continuously differentiable).
  This part of the
energy depends on the parameters
$\tilde{\rho}_c, \tilde{\lambda}_c, c, k_4, k_5, a$, which have the
following interpretations.  For $\lambda >c$ the energy contribution
to $\psi_2$ vanishes (i.e., the bond is broken). The values
$\tilde{\rho}_c, \tilde{\lambda}_c$ are the minimizing
values of $\rho, \lambda$ for $\psi_2$; in practice, we adjust
these so that the measured values ${\rho}_c, {\lambda}_c$
are absolute minimizers
of the total energy.
The value $a$ is the bond dissociation
energy; $k_4$ controls the stiffness of this bond with respect to
changes of radius, and $k_5$ favors uniformity.   The term
containing $k_5$  multiples $\calL$ so that the tendency
toward uniformity is not in force
when the bond is broken.
 The third bond  $(i,j),(i-1,j+2)$ is similar
to the second, in that it forms upon contraction.
It spans two helices, so the third
bond energy depends on the pitch and radius of the
second nearest helix, and it involves the larger set
of variables $\rho_j, \rho_{j+1}, \rho_{j+2}, \bar{\gamma}_j,
\bar{\gamma}_{j+1}$.  We take it to have a simple
form similar to that of the second bond,
\beqs
\lefteqn{\psi_3(\y_{i,j},\R_{i,j} ,\y_{i-1,j+1},\R_{i-1,j+1})}
\nonumber \\
& & = \frac{1}{12}\left( 1-k_4(\rho_j-\tilde{\rho}_c)^2
-k_5 (\rho_{j+1}- \rho_j)^2
-k_6 (\rho_{j+2}- \rho_j)^2
-k_7 (\bar{\lambda}_{j+1}- \bar{\lambda}_j)^2\right)
{\calL}(\bar{\lambda}_j),    \nonumber
\eeqs
where $\calL$ is as in (\ref{calL}). In principle, all of the
parameters $\tilde{\rho}_c, \tilde{\lambda}_c, c, k_4,k_5, a$ are
likely to differ for bonds 2 and 3, but we do not alter the
notation to reflect that.

So, in summary, for the constrained sheet subject to the 1-D
ansatz, we write the total free
energy
\beq
\Psi(\y_{1,1},\Rvec_{1,1}\, , \ldots,\, \y_{6, 23},
\R_{6, 23}) = \sum_{ j \in \{1, \dots, 23 \} }
\psi(\rho_j, \rho_{j+1}, \rho_{j+2}, \bar{\lambda}_j,
\bar{\lambda}_{j+1})  \label{gen1D}
\eeq
where the energy per annulus
$\psi = 6\, ( \psi_1 + \psi_2 + \psi_3)$ and
$k_i>0, i=1, \dots, 7, \ a>0,\ k_1k_2-k^2 >0,
\ \lambda_c < c < \lambda_e $.
 Note that, because of presence of the cut-off,
the values $\rho_e, \lambda_e$ are always relative
minimizers of the energy if (as we assume)
the stiffness matrix in (\ref{bond1})
is positive-definite.

This energy favors uniform configurations for a suitably
restricted domain and for ranges of the
parameters expected to be physically interesting.  Consider
the domain $(\lambda_j, \rho_j)$  where $\calL<0$ and the
prefactor of $\calL$ is positive.   Then a lower
bound for the energy on this domain
is obtained by putting $k_3 = k_5 = k_6 = 0$ and this
bound is achieved by a uniform configuration that minimizes
each term (The individual terms of the sum are minimized at
the same uniform state).  We use the notation
\beq
\phi(\rho,\lambda) =
\psi(\rho, \rho, \rho, {\lambda}, {\lambda})
\label{prl}
\eeq
for the energy per annulus of uniform states.

A simple explicit energy that uses all of the measured data that
we have available, but otherwise makes somewhat arbitrary choices
of constants, and has a relative minimizer at the
extended state and an absolute minimizer at the contracted
state, is obtained by putting
$k_1 = 0.333\ {\rm zcal}\ {\text \AA}^{-2},\
 k_2  = 3.0 \ {\rm zcal}\ {\text \AA}^{-2}
,\  k = 0,\
k_4 = 10^{-4}\ {\text \AA}^{-2},
\ a = 0.3719\ {\rm zcal}\ {\text \AA}^{-3},\
c = 30\ {\text \AA},\ \tilde{\lambda}_c = 13.3901\ {\text \AA},
\ \tilde{\rho}_c = 161.398\ {\text \AA}$, and then by evaluating at a
uniform state (1 zcal = $10^{-24}$ kcal).  This gives the double-well energy pictured in
Figure \ref{special}.  Here the choice of $a$ reflects the
calorimetric measurement of Arisaka, Engel and Klump \cite{{aek_81}}
that gives
$ \phi(\rho_e,\lambda_e)-\phi(\rho_c,\lambda_c) =
60$ zcal/annulus, based on arguments described at the
end of this subsection.

\begin{figure}
\includegraphics[width=.9\columnwidth]{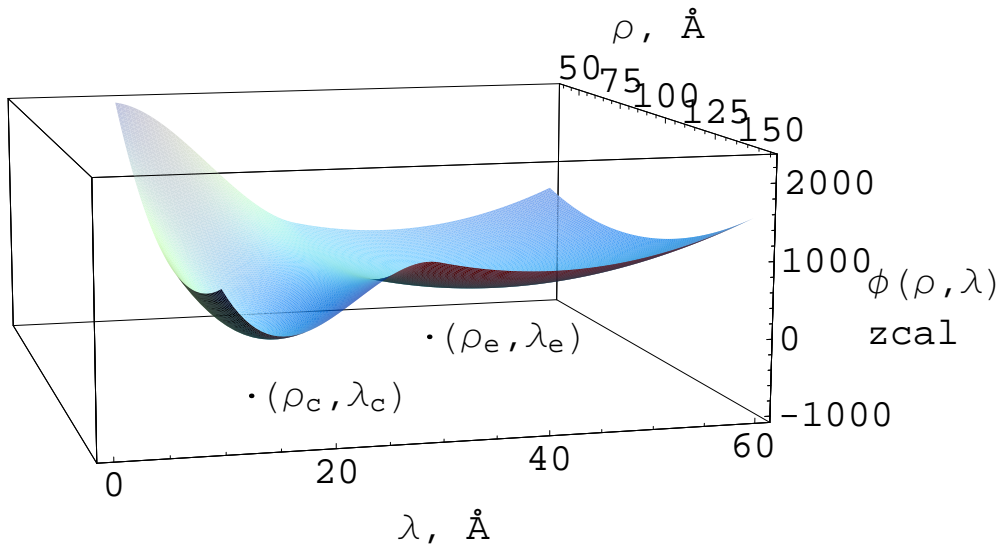}
\caption{Special energy for tail sheath. \label{special}}
\end{figure}

The tendency toward uniform states plays an important
role during self-assembly of tail sheath.
During assembly, the baseplate forces the first annulus
to have the extended radius,  $\rho_1 = \rho_e$.  As
subsequent annuli are added they
do so as to match the radius of the annulus below.  The
second and third
bonding directions remain incomplete, since the formation
of these bonds would require
the annuli to adopt the contracted radius.
Proper assembly is accomplished by design; the penalty
for mismatching a neighboring annulus
outweighs the energy the could be liberated by forming
the additional bonds. Our energy given above has the flexibility
to model this behavior through two features: 1) the state
$(\rho_e, \lambda_e)$  is a minimizer of energy with respect
to all uniform small perturbations of $(\rho, \lambda)$, and
2) the terms involving $k_3, k_5, k_6, k_7$ can be tuned so
that the addition of a new layer onto the growing extended
sheet would be penalized from being added with $(\rho, \lambda)$
near the contracted values, even though these have lower
uniform energy.  The complete analysis of
self-assembly would require a molecule-by-molecule growth
mechanism, involving boundary energies, but the present
energetic considerations are expected to play a role.

For the rest of the paper we use a more general energy
than the special form given above, but one that retains
some of its essential features.  That is we assume an energy
per annulus of the form $\psi(\rho_j, \rho_{j+1}, \rho_{j+2},
\bar{\lambda}_j, \bar{\lambda}_{j+1})$
(cf., (\ref{gen1D})) having the properties
\beqs
\psi(\rho_c, \rho_c, \rho_c, \lambda_c, \lambda_c) &<&
\psi(\rho_1, \rho_2, \rho_3, \lambda_1, \lambda_2)  \ \ \
\left\{ \begin{array}{l}
{\rm for\ all}\ \  (\rho_1, \rho_2, \rho_3, \lambda_1, \lambda_2) \\
{\rm not\ equal\ to}\ \ (\rho_c, \rho_c, \rho_c, \lambda_c, \lambda_c),
\end{array} \right.
\nonumber \\
\psi(\rho_e, \rho_e, \rho_e, \lambda_e, \lambda_e) &<&
\psi(\rho_1, \rho_2, \rho_3, \lambda_1, \lambda_2)  \ \ \
\left\{ \begin{array}{l}
{\rm for\ all}\ \  (\rho_1, \rho_2, \rho_3, \lambda_1, \lambda_2) \\
{\rm near,\ but\ unequal\ to,} \\
 (\rho_e, \rho_e, \rho_e, \lambda_e, \lambda_e),
\end{array} \right.
\nonumber \\
\psi(\rho_c, \rho_c, \rho_c, \lambda_c, \lambda_c) &<&
\psi(\rho_e, \rho_e, \rho_e, \lambda_e, \lambda_e).
\label{psiwells}
\eeqs
For this more general energy we retain the notation
$\phi(\rho,\lambda) =
\psi(\rho, \rho, \rho, {\lambda}, {\lambda})$, so it
follows from the above that
\beqs
 \phi(\rho_c, \gamma_c) &<&  \phi(\rho, \gamma)
 \ \ \ {\rm for\ all}  \ \  (\rho, \gamma) \ne (\rho_c, \gamma_c),
 \nonumber \\
  \phi(\rho_e, \gamma_e)  &<&  \phi(\rho, \gamma)
 \ \ \ {\rm for\ all}  \ \  (\rho, \gamma) \ne (\rho_e, \gamma_e)
 \   {\rm but\ near}
 \  (\rho_e, \gamma_e).  \label{phi_1}
\eeqs

We make one other assumption on the height difference between
the energy wells.  In
\cite{aek_81} Arisaka et al. did calorimetry on T4 tail
sheaths with contraction
triggered by two methods: raising the temperature to 72 C and by
increasing the concentration of urea.  The former gave
-44 kcal/mole (of gp18 molecules) whereas the latter gave
-25 kcal/mole (of gp18 molecules).  We use the former number here
as it was considered the more accurate by these authors.
From the details of the measurement,
raising the temperature did not give reversible contraction, but
rather irreversible contraction, and temperatures higher than
72 C caused denaturation of the whole sheath.  Thus, one can infer
that the free energy of contracted sheath is still lower than
of extended sheath at 72 C, though not as low as at 25 C.  Without
any additional information and considering that at least
spontaneous contraction occurred at 72 C, we estimate the
height difference between the wells by the following procedure.
We restore the temperature dependence of $\phi$ and  Taylor expand
in the temperature,
omitting the error terms,
\beq
 \phi(\rho, \lambda, \theta_2) =  \phi(\rho, \lambda, \theta_1)
 + \frac{\partial \phi(\rho, \lambda, \theta_1)}{\partial \theta}
 (\theta_2-\theta_1).  \label{Te}
\eeq
We put $\theta_1  = 25$ C and $\theta_2  = 72$ C,
evaluate (\ref{Te}) at $(\rho_e, \lambda_e)$ and
$(\rho_c, \lambda_c)$  and subtract, estimating
$ \phi(\rho_e, \lambda_e, \theta_2) \approx
\phi(\rho_c, \lambda_c, \theta_2) $.
 Now, as is common in the interpretation
of calorimetric measurements of phase transformations, we
interpret the temperature times entropy difference as
the latent heat:
\beq
 \theta_2 (\frac{\partial \phi(\rho_c, \lambda_c, \theta_1)}
 {\partial \theta} - \frac{\partial \phi(\rho_e, \lambda_e, \theta_1)}
 {\partial \theta}) = 440\ \  {\rm zcal/annulus}.  \label{Lh}
\eeq
Here we have ignored the temperature dependence of the
entropy evaluated at either well separately.   Combining
(\ref{Te}) and (\ref{Lh}) we get that the entropy difference
of the two phases is $1.27$ (zcal/K annulus) and that
\beq
 \phi(\rho_e,\lambda_e)-\phi(\rho_c,\lambda_c) =
60 \  {\rm zcal/annulus}. \label{60}
\eeq

\subsection{Some simple uniform deformations and some relations
between moduli}

For the purpose of defining various moduli,
it is convenient to introduce the free energy per unit
reference length (the reference being the contracted state)
by defining
\beq
  \phi_c(\rho,\gamma) = \frac{1}{ \lambda_c}\, \phi(\rho,\lambda).
  \label{phi_c}
\eeq
Second derivatives of $\phi_c(\rho, \lambda)$ with
respect to the pair $(\rho, \lambda)$ have interpretations
as various moduli. For example, if we consider small deformations
about, say, contracted sheath, then we  write
\beq
\phi_c(\rho, \lambda) =
\phi_c^0 + \frac{1}{2} \left(A (\rho-\rho_c)^2 +
2 B (\rho-\rho_c) (\lambda-\lambda_c)
+ C (\lambda-\lambda_c)^2 \right)+ \dots,  \label{philin}
\eeq
where $\phi_c^0$ is the free energy per unit length of
undistorted contracted
sheath.  We assume this form is positive-definite.

We now interpret these moduli $A, B, C$.  Working within the 1D
ansatz, suppose we hold the annulus $j = 1$ fixed and
apply an axial force $\fvec = f \evec_3$ to annulus $j = 23$,
treated as a dead load.
Then the total energy of sheath and loading device is
\beq
 \Psi(\y_{1,1},\Rvec_{1,1}\, , \ldots,\, \y_{6, 23}, \R_{6, 23})
 - \y_{1,23} \cdot \fvec =
 \Psi(\y_{1,1},\Rvec_{1,1}\, , \ldots,\, \y_{6, 23}, \R_{6, 23})
 - 22\, f\,  \lambda_{23}    \label{fen}
\eeq
(Recall the relation between $\lambda_j$ and $\bar{\lambda}_j$,
equation (\ref{inversenewvars})).  Using the assumptions
(\ref{psiwells}) and the argument just preceding (\ref{prl}),
 we see that the minimizing state $\{ \rho_1, \dots, \rho_{23} \}$,
$\{ \bar{\lambda}_1, \dots, \bar{\lambda}_{23} \}$ is uniform,
\beq
  \rho_1 = \dots = \rho_{23}  = \rho, \ \ \ \
  \bar{\lambda}_1=  \dots =  \bar{\lambda}_{23}  = \lambda,
\eeq
and $(\rho,\lambda)$ minimizes
\beq
\phi_c(\rho, \lambda) -  f \,
\frac{\lambda}{\lambda_c}.   \label{tenmin}
\eeq
Minimizing this expression over $(\rho, \lambda)$ for small
values of $f$, we get
\beqs
\lambda - \lambda_c &=&
\frac{ A }
{\lambda_c(AC-B^2)}\ f + \dots,  \nonumber \\
 \rho - \rho_c  &=&
\frac{ -B }
{\lambda_c(AC-B^2)}\ f + \dots,   \nonumber \\
\gamma - \gamma_c &=&
\frac{-A}{\rho_{eff}^2 \gamma_c (AC-B^2)}\ f + \dots.
\eeqs
The tensile modulus (i.e., the proportionality factor between
$f$ and $(22 \lambda - 22 \lambda_c)/ 22 \lambda_c$) is
therefore
\beq
   {\rm tensile\ modulus} = \frac{\lambda_c^2 (AC-B^2)}{A}\, .
   \label{tense}
\eeq
Hence, due to the positive-definiteness of the quadratic form
(\ref{philin}), tensile force produces extension, and also
twist, with an end angle that decreases with increasing force.
We expect $B>0$ in which case
the Poisson effect is the usual one: lengthening produces
a decrease in the radius. We can define a ``Poisson's ratio''
via the usual formula ($-$radial strain/axial strain):
\beq
{\rm Poisson's\ ratio} = \frac{\lambda_c}{\rho_c}
\frac{B}{A}
\eeq

For simple torsion defined by the loading device energy
$- 22 M \gamma $, where $M \evec_3$ is the applied moment,
energy minimization of $\phi_c - M (\gamma/\lambda_c)$,
analogously to the above, leads to uniformity and
to the equations
\beqs
\gamma - \gamma_c &=& \frac{ A\, \lambda_c }
{\gamma_c^2\, \rho_{eff}^4\, (AC - B^2)}\ M + \dots,   \nonumber \\
 \rho - \rho_c  &=& \frac{ B }
{\rho_{eff}^2\, \gamma_c \, (AC-B^2)}\ M + \dots,  \nonumber \\
 \lambda - \lambda_c  &=& \frac{ -A }
{\rho_{eff}^2\, \gamma_c \, (AC-B^2)}\ M + \dots\ .
\eeqs
From here, we identify
\beq
{\rm torsional\ modulus} = \frac{\rm moment}{\rm twist/length} =
\frac{\gamma_c^2\, \rho_{eff}^4 (AC-B^2)}{A}    \label{torsi}
\eeq
Thus we predict that the
torsional modulus is  proportional to the tensile modulus,
the proportionality factor only depending on the geometry
of contracted sheath. In contrast, in macroscopic
elasticity the torsional and tensile moduli are governed
by different elastic constants (i.e., the shear and Young's
moduli, respectively).  This unusual behavior arises
from our unusual constraints.

Finally, we briefly consider the resistance of protein
structures to internal pressure.
This may be relevant to the interactions between the sheath
and tail tube, so in fact it is more related to extended
sheath (It is also of course highly relevant to the packaging of
DNA in capsids \cite{phillips}).  For internal pressure $p$, the
associated loading device energy is $23 \lambda \pi \rho^ 2 p$.
It is trivial to work out the associated moduli so
we do not record that here.

What is more interesting is to work out the reaction forces.
When a body is constrained, there should be reaction forces,
that is, certain kinds of forces that do not produce deformation.
In the present setting but in the fully nonlinear case,
we consider a sheath subject to a
tensile force $f$, a twisting moment $M$ and an internal pressure
$p$ altogether, with an associated loading device energy
$-22 f \lambda - 22 M \gamma + 22 \lambda \pi \rho^2 p$. Taking
the first variation of the energy with respect to $(\rho, \gamma)$
we see that the resulting two equations do not uniquely determine
the three unknowns $(f,M,p)$. In fact it turns out that changes
of $p$ generically lead to deformation and the constraint force
only involves $f$ and $M$.   The result can be stated in the
following way.  Suppose that the sheath is in equilibrium at
a state $(\tilde{\rho}, \tilde{\gamma})$ corresponding to
generalized forces $(f,M,p)$.  Then,
$(\tilde{\rho}, \tilde{\gamma})$ is in
equilibrium if $(f,M)$ are changed to $(f + f_1, M + M_1)$, where,
\beq
 M_1 = -\frac{d \lambda}{d\gamma}\ f_1 = \frac{\rho_{eff}^2 \tilde{\gamma}}
 {\tilde{\lambda}}\  f_1 .  \label{reaction}
\eeq
Alternatively, this condition can be thought of in terms of work:
changes of force and moment consistent with (\ref{reaction})
do no work on the sheath.

\section{Relaxed states} \label{sect9}

It is interesting to contrast our theory of the protein sheet with
nonlinear continuum theories of plates and shells.  As a related example,
the mechanical behavior of carbon nanotubes have been shown to
conform to such continuum theories in many aspects,
especially regarding elasticity and buckling \cite{ab_02}.
The results given above arising from the
constraints, especially the first order Poynting effect and the
relations between elastic moduli, suggest differently, but these
results are closely connected with the presence of the constraints.
As we show here, the predictions of the unconstrained theory are
also essentially different
from continuum theories.  This is not fundamentally a ``nanoscale''
phenomenon, but is related to the particular structure
of protein sheet: compact globular proteins with local bonding
and a sensitivity to orientation.
 
To review, nonlinear continuum theories of thin plates and shells
come in various varieties, depending on the strength of the
applied forces (for a rigorous treatment and an overview of
the regimes, see \cite{fjm_05}). For the largest applied
forces there is membrane
theory, defined in the following way: let $\vphi(\Gvec)$ be the
three-dimensional
nonlinear elastic energy of the material expressed as a function
of the deformation gradient $\Gvec$.  We suppose as usual
that $\vphi$ is frame-indifferent,
$\vphi(\Qvec\Gvec) = \vphi(\Gvec)$ for all $\Qvec \in\ $SO(3)
and that  $\vphi$ is minimized on SO(3).
Let
$(\evec_1,\evec_2,\evec_3)$
be an orthonormal basis with $\evec_3$ normal to the plate in its
reference configuration, and write
$\Gvec = G_{ij}$ in this basis. To describe membrane theory we
express the deformation
gradient as its three column vectors,
$\Gvec = (\yvec_1 | \yvec_2 | \yvec_3)$.  If $\yvec(x_1,x_2),
\ (x_1,x_2) \in S$ is
the deformation of the plate then, in the absence of a loading
device, the energy is
\beq
 \int_S \vphi (\yvec,_1| \yvec,_2| \bvec) \, dx_1 dx_2.
 \label{membrane}
\eeq
This is minimized over the independent fields
$(\yvec(x_1,x_2), \bvec(x_1,x_2))\ $ ($\bvec$ describes deformations
relative to the ``middle surface'').  Suppose that we have no
boundary conditions imposed.  Then, the energy in
(\ref{membrane}) is minimized by $(\yvec,\bvec)$  satisfying
\beq
 \yvec,_1 = \Rvec(x_1,x_2) \evec_1, \
 \yvec,_2 = \Rvec(x_1,x_2) \evec_2, \
 \bvec = \Rvec(x_1,x_2) \evec_3,  \ \ \
  \Rvec(x_1,x_2) \in {\rm SO(3)}.  \label{isometric}
\eeq
The third of these equations simply determines $\bvec$, while
the first two restrict the deformation.  In fact, the first
two of these equations define so-called {\it isometric
mappings}.  Isometric mappings are essentially the mappings that
one can illustrate by taking a flat sheet of paper and deforming it,
including the possibility of making folds and rather complex
``crumpling''.   At the next level of
approximation, for weaker forces,  we have nonlinear bending theory.
This is defined by
the same kinematics as just described, but with the energy
\beq
 \frac{1}{24}\int_S
 \psi((\nabla \yvec)^T \nabla \bvec) \, dx_1 dx_2,
 \label{membrane2}
\eeq
where $\psi(\Gvec)  = \min_{\cvec} q(G_{ij} + c_i\ \delta_{3 j})$,
$\delta_{ij}$ is the Kronecker delta,
and $q$ is the quadratic form
\beq
 q(\Hvec) =
 \frac{\partial^2 \vphi(\Ivec)}
 {\partial F_{ij}\partial H_{km}} H_{ij} H_{km}.  \label{bending}
\eeq
In this case the energy (\ref{bending}) is minimized
over isometric mappings {\it only},
i.e. over the solutions of (\ref{isometric}).  Thus, in summary,
isometric mappings are the basic relaxed states of plate
theories: they are the zero energy deformations of membrane theory
and the finite energy deformations of bending theory. Shell
theories are variants of the above in which the given
reference state is generally curved.  In that case the
finite energy deformations are isometric mappings of the curved
reference state.

What are the relaxed states of the present theory?  To calculate the
analog of the above, we should minimize the energy of the sheet
\beqs
\lefteqn{\Psi(\y_{1,1},\Rvec_{1,1}\, , \ldots,\, \y_{N, M}, \R_{N, M})}
\\ \nonumber
&=& \sum_{(i,j)\, \in\, \ZZ^2\, \cap\, {\cal D} }
\psi_1(\y_{i,j},\R_{i,j},\y_{i+1,j},\R_{i+1,j}) +
\psi_2(\y_{i,j},\R_{i,j} ,\y_{i,j+1},\R_{i,j+1}),
\label{Psisheet1}
\eeqs
without loading device energies or boundary conditions, over
all positions and orientations.  Here we have assumed two bonding
directions and we have ignored boundary molecules.  To simplify,
this can be written (modulo possibly a few missing or additional
boundary molecules)
\beq
\Psi = \sum_{(i,j)\, \in\, \ZZ^2\, \cap\, {\cal D} }
\tilde{\psi}_1(\tvec_{i,j},\Qvec_{i,j}) +
\tilde{\psi}_2(\hat{\tvec}_{i,j},\hat{\Qvec}_{i,j}),
\label{Psisheet2}
\eeq
but now it must be born in mind that the independent variables
$(\tvec_{i,j} ,\Qvec_{i,j}, \hat{\tvec}_{i,j} ,\hat{\Qvec}_{i,j})$
are subject to the compatibility conditions (\ref{com}). These
conditions, repeated here
\beqs
\hat{\Qvec}_{i,j} \Qvec_{i,j+1}\hat{\Qvec}^T_{i+1,j}\Qvec^T_{i,j} &=& \Ivec,
\nonumber \\
 \hat{\tvec}_{i,j} + \hat{\Qvec}_{i,j} {\tvec}_{i,j+1}
 - \Qvec_{i,j} \hat{\tvec}_{i+1,j} -  \tvec_{i,j} &=&  0,
 \label{com1}
\eeqs
couple molecules $(i,j)$ with
$(i+1,j)$ and $(i,j+1)$.  Clearly, we cannot (as we did above)
minimize
(\ref{Psisheet1}) by minimizing the ``integrand''
$\tilde{\psi}_1(\tvec_{i,j},\Qvec_{i,j}) +
\tilde{\psi}_2(\hat{\tvec}_{i,j},\hat{\Qvec}_{i,j})$, for
this would typically give a minimizer, say, of the form
\beq
\left.
\begin{array} {l}
 \tvec_{i,j} = \tvec, \\
 \Qvec_{i,j} = \Qvec, \\
 \hat{\tvec}_{i,j} = \hat{\tvec}, \\
 \hat{\Qvec}_{i,j} = \hat{\Qvec},
 \end{array} \right\}
 \quad {\rm where} \quad
\left\{\begin{array} {l}
 \tilde{\psi}_1(\tvec,\Qvec) \le
 \tilde{\psi}_1(\avec ,\Rvec)\ \
  {\rm for\ all}\ \  (\avec ,\Rvec), \\
     \\
 \tilde{\psi}_2(\hat{\tvec},\hat{\Qvec})   \le
 \tilde{\psi}_2(\hat{\avec} ,\hat{\Rvec})\ \
 {\rm for\ all}\ \  (\hat{\avec} ,\hat{\Rvec}),  \\
 \end{array} \right.
\eeq
and it is seen that such a minimizer would generically
fail the compatibility conditions, which in this case
become
\beqs
\hat{\Qvec}\Qvec \hat{\Qvec}^T\Qvec^T
 &=& \Ivec,
\nonumber \\
 \hat{\tvec} + \hat{\Qvec} {\tvec}
 - \Qvec \hat{\tvec} -  \tvec &=&  0.
 \label{com2}
\eeqs
One can consider more complex rearrangements of the sum
(\ref{Psisheet1}), with different ``integrands'',
 but the analogous problem arises again.
For example, the apparently most promising rearrangement is
the sum\footnote{We ignore possible problems with this rearrangement
near the boundary.}
\beq
\Psi = \sum_{\begin{array}{c}(i,j)\, \in\, \ZZ^2\, \cap\, {\cal D} \\
 i+j = {\rm even} \end{array} }
\tilde{\psi}_1(\tvec_{i,j},\Qvec_{i,j}) +
\tilde{\psi}_2(\hat{\tvec}_{i,j},\hat{\Qvec}_{i,j}) +
\tilde{\psi}_1(\tvec_{i,j+1},\Qvec_{i,j+1}) +
\tilde{\psi}_2(\hat{\tvec}_{i+1,j},\hat{\Qvec}_{i+1,j}).
\label{Psisheet3}
\eeq
Here the summand contains exactly the independent variables
appearing in the constraints, and therefore we
could minimize it with respect to all values of the
independent variables subject to the constraints. But one
then sees that the solution actually satisfies only the compatibility
conditions on every other cell and generically does not give a
minimizer.

In discrete theory the impossibility of minimizing the energy
for each each bond individually is termed {\it frustration}.
In continuum theory the concept is similar \cite{sethna}: it is
the inability of minimizers of the
energy density to satisfy conditions
of compatibility inherent in the kinematics\footnote{A simple
example of frustration in continuum
theory is illustrated
by an energy $\int_{{\cal D}} |\nabla z - \Avec \xvec|^2
d\xvec$, where $\Avec$ is a nonsymmetric matrix.  In this case there
is a unique smooth minimizer but it is not obtained by minimizing the
integrand.}.  We
can say that our sheet is also frustrated, in the sense that
minimization of the energy density for each bond does not
generically give a compatible deformation.  Here, the word
``generically'' means that, even if these compatibility conditions
happen to be (accidentally) satisfied for a minimizer, then they
are not satisfied if $\tilde{\psi_1}$ or $\tilde{\psi_2}$  are
smoothly perturbed consistent with all of their assumed
symmetries. As indicated above, even if we allow small collections
of multiple bonds and minimize the energy of these, subject to
constraints of compatibility, we also obtain a configuration that is
not compatible in the large.

In biology, unlike materials science, there is the phenomenon
of evolution of materials to achieve fitness.  Thus for a
protein sheet, there might be reasons, for
example, to achieve a particularly low energy
state, for a protein sheet to be nongeneric. Thus it is of
interest to assume that  ${\tvec}, {\Qvec}$   and
$\hat{\tvec}, \hat{\Qvec}$ minimize, respectively,
$\tilde{\psi}_1, \tilde{\psi}_2$ and also satisfy (\ref{com2})
and then to see what kinds of sheets emerge. We
call such states {\it fully relaxed states}: each bond is relaxed
and the configuration is compatible.

To calculate all fully  relaxed states, we merely have to characterize
all solutions of (\ref{com2}) and then calculate the implied
positions and orientations.  This is a straightforward algebraic
exercise and we just give the results.  First, a useful
characterization of the solutions of (\ref{com2}) is the following.
Each solution falls into one of the categories below:
\begin{enumerate}
\item $\Qvec$ and $\hat{\Qvec}$ are coaxial,
$\Qvec \evec = \hat{\Qvec} \evec = \evec, |\evec| = 1$, and
\begin{enumerate}
\item If $\Qvec \ne \Ivec$ and $\hat{\Qvec} \ne \Ivec$, then
$\tvec = \tvec_1 + \tau \evec$ and $\hat{\tvec} =
\hat{\tvec}_1 + \hat{\tau} \evec$ for some $\tau, \hat{\tau}$
with ${\tvec}_1 \cdot \evec =
\hat{\tvec}_1 \cdot \evec = 0$, and $\hat{\tvec}_1  =
(\Ivec - \Qvec)^{-1}(\Ivec - \hat{\Qvec}){\tvec}_1 $, the inverse
taken on the plane perpendicular to $\evec$.
\item If $\Qvec \ne \Ivec$ and $\hat{\Qvec} = \Ivec$, then
${\tvec}$ is arbitrary but  $\hat{\tvec} = \tau \evec$ for some $\tau$.
\item If $\Qvec = \Ivec$ and $\hat{\Qvec} \ne \Ivec$, then
$\hat{\tvec}$ is arbitrary but  ${\tvec} = \tau \evec$ for some $\tau$.
\item If $\Qvec = \Ivec$ and $\hat{\Qvec} = \Ivec$, then
$\hat{\tvec}$ and   ${\tvec} $ are arbitrary.
\end{enumerate}
\item $\Qvec = -\Ivec + 2 \evec \otimes \evec$
and $\hat{\Qvec} =  -\Ivec + 2 \hat{\evec} \otimes \hat{\evec}$,
$|\evec| =  |\hat{\evec}| = 1, \evec \cdot \hat{\evec} = 0$, and
\begin{enumerate}
\item $\hat{\tvec} = \tau_1 \evec + \tau (\evec \times \hat{\evec})$,
$\tvec =  \tau_2 \hat{\evec} + \tau (\evec \times \hat{\evec})$ for
some $\tau_1, \tau_2, \tau$.
\end{enumerate}
\end{enumerate}

Now, using these results we go back and compute the uniquely
determined (up to overall rigid body motion) positions and
orientations. For simplicity we assume an $N \times M$ sheet.
 We find, in all cases,
\beqs
\Rvec_{i+1,j+1} &=& \Rvec_{1,1}\Qvec^{i} \hat{\Qvec}^{j},
\nonumber \\
\yvec_{i+1,j+1} &=&  \yvec_{1,1}  +
\Rvec_{1,1}\left[ \sum_{k = 0}^{j-1}\hat{\Qvec}^k \hat{\tvec}
+ \hat{\Qvec}^{j}\sum_{k=0}^{i-1}\Qvec^k \tvec \right],
\nonumber \\
& & i = 1, \dots, N,  \ \ \ j = 1, \dots, M.
\label{fr}
\eeqs
But this is exactly of the form of (\ref{T4sheath})
for T4 tail sheath!  (Note: recall that in the formula
for tail sheath the $\Rvec_{1,1}$ was moved
through the $\Qvec^{i} \hat{\Qvec}^{j}$ using the remark
given at the end of Section \ref{sec5})).

It is intriguing to ponder whether these fully relaxed states
are actually realized by tail sheath (or other protein
sheets) and, if so, the implications
of this with regard to stability, and evolutionary development.
 Of course, with the various choices of
 ${\tvec}, {\Qvec}, \hat{\tvec}, \hat{\Qvec}$  as enumerated above,
 the sheet will not look exactly like tail sheath.  We explored
 this numerically by choosing various cases and found that the
 general appearance is however much like tail sheath;  in fact, it can be
 proved from the formula  (\ref{fr}) that the bonding direction
 $(i,j)-(i-1,j+1)$ is also a helix\footnote{Apparently, the formula
 produces helical configurations in all rational directions of $\ZZ^2$.}
 (with translation
 $\Qvec(\hat{\tvec} + \tvec)$). Figure \ref{figure10} shows a
 generic picture of a fully relaxed state, with the molecule
 represented by a simple ellipsoid.

To complete this story, we make brief remarks about the
remaining cases
 of the enumeration above.  In item 1, if
either $\Qvec = \Ivec$ or $\hat{\Qvec} = \Ivec$ then the
appearance is still more or less like Figure \ref{figure10},
but one family of helices degenerates to
straight lines of molecules that are parallel to
the axis of the cylinder. If both $\Qvec = \Ivec$ and
$\hat{\Qvec} = \Ivec$, then the cylinder degenerates to a
planar sheet, with crystalline symmetry, and parallel orientations
of molecules.  Finally, item 2 is a bit surprising;  it describes
a collection of four molecules, not generally at the corners of
a regular tetrahedron, but such that each pair of the molecules
is twinned, that is, individuals of the pair are
related by a 180$^\circ$ rotation.

In summary, our theory of a protein sheet is generically
frustrated.
Energy minimizers are generally naturally curved,
as in  shell theories,
but this curvature is determined by the energy.
Isometric mappings seem to play no role here.   In our
theory if one considers energy densities $\tilde{\psi}_1,
\tilde{\psi}_2$  that are minimized
at compatible pairs $(\tvec, \Qvec), (\hat{\tvec}, \hat{\Qvec})$
(i.e., fully relaxed states) then the energy minimizers
look much like the tail sheath of bacteriophage T4 and are given by
simple formulas.
\begin{figure}
\includegraphics[width=.3\columnwidth]{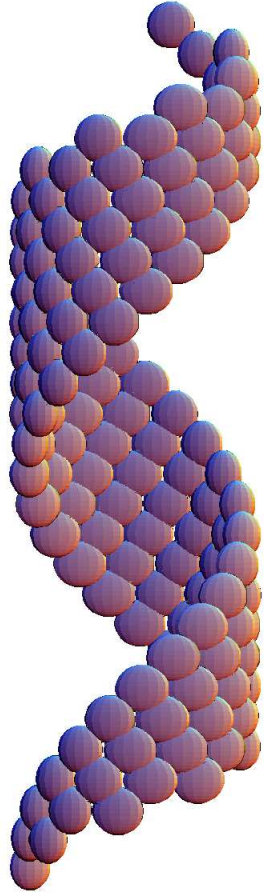}
\caption{A generic fully relaxed state. }
\label{figure10}
\end{figure}

\section{Experiments suggested by the theory}

We have noted above several places above where there are
possible experimental tests of our predictions.  These include
the extension-twist relation (Figure \ref{figure6}), the
linearized behavior near contracted or extended sheath
(\ref{tense})-(\ref{torsi}), and the reaction
forces that preserve deformation (\ref{reaction}).
We now discuss two other types of predictions that relate
directly to biological and interesting nonbiological behavior.

\subsection{The force of penetration}
One of the most important predictions of our model is the
force of penetration.
Consider applying an axial force $f$ to contracted sheath, so as
to stress-induce the transformation to extended sheath.
Alternatively, one can imagine applying sufficient tension to
extended sheath to just prevent contraction, i.e., the stall
force.  We neglect
the interactions with the tail tube,
assuming it to be weakly
bonded to the sheath even when it is in the extended state,
as is thought to be true, \cite{watts}, \cite{leiman1}.
For small
values of $f$ the behavior is given by the analysis of
(\ref{fen})-(\ref{tense}), and we expect the  initial slope of
the force-elongation curve ($f$ vs.
$(\lambda - \lambda_c)/\lambda_c$) given by the modulus
${\lambda_c^2 (AC-B^2)}/{A}$.
There is expected to be significant
nonlinearity of the response, because the constraints themselves
are nonlinear.  The details of the response near transition may
depend on details of the loading device -- whether hard or
soft, for example -- but one expects some kind of load drop
on nucleation.  The transformation is expected to take place
via movement of an interface, as pictured qualitatively
in Figure \ref{figure8} and at approximately
constant free energy, because a bias of free energy toward
either phase would, by energy minimization, tend to drive out
the interface, one way or the other.  This suggests the criterion
\beq
\Psi(\y^c, \Rvec^c) - 22\,  f\,  \lambda^c_{23} =
\Psi(\y^e, \Rvec^e) - 22\,  f\,  \lambda^e_{23},  \label{efeten1}
\eeq
where the uniform states $(\y^{c,\, e}, \Rvec^{c,\, e}) =
(\y^{c,\, e}_{1,1},\Rvec^{c,\, e}_{1,1}\, , \ldots,\,
\y^{c,\, e}_{6, 23}, \R^{c,\, e}_{6, 23})$  are assumed to
be in equilibrium.   Let superscripted variables
$(\lambda^{e,\,c}, \rho^{e,\,c}, \gamma^{e,\,c})$ be associated with these
uniform equilibrium states.
Using the special form of the energy  (\ref{phi_1})-(\ref{phi_c}),
we get
\beq
\phi_c(\rho^c, \lambda^c) -  f \,
(\lambda^c/\lambda_c) =  \phi_c(\rho^e, \lambda^e) -  f \,
(\lambda^e/\lambda_c),    \label{efeten2}
\eeq
where
\beq
\frac{\partial \phi_c(\rho^{c,\, e}, \lambda^{c,\, e})}{\partial \rho}
 = 0, \ \ \ \ \
 \frac{\partial \phi_c(\rho^{c,\, e}, \lambda^{c,\, e})}
 {\partial \lambda}  =    \frac{f}{\lambda_c}.    \label{efeten3}
\eeq
Solving these together,  we get,
\beqs
f &=& f_{trans} =  \
\frac{(\phi_c(\rho^e, \lambda^e) - \phi_c(\rho^c, \lambda^c))\lambda_c}
{ (\lambda^e - \lambda^c)} \nonumber \\
 &=&
\frac{(\phi_c(\rho_e, \lambda_e) - \phi_c(\rho_c, \lambda_c))}
{(\frac{\lambda_e}{\lambda_c} - 1)} +
{\rm O}\left(\max_{c,e}\frac{A_{c,\, e}}
{A_{c,\, e}C_{c,\, e} - B_{c,\, e}^2} f^2   \right). \label{efeten4}
\eeqs
The error term depends also on geometric factors and can be
written explicitly, but we note that it is of the form
($f^2/$tensile modulus).  Thus, if $f \ll $ tensile modulus
then this term is negligible as compared to $f$, and the
force at transformation is a simple ratio of the height difference
between the energy well minima at extended
and contracted sheath and the
difference between the lengths of the sheath. These kinds
of results are well known in the study of phase transformations.

Of course, the virus uses the reverse transformation, from
extended to contracted sheath, during penetration.  The maximum
force available for penetration is expected to be also $f_{trans}$.

We can evaluate the force of contraction based on the height
difference between the wells, accounting for the reservations
given at the end of Subsection \ref{sect8.3}.  Combining
(\ref{efeten4}) and (\ref{60}), we get,
\beq
 f_{trans} = 103\  {\rm  pN}
\eeq
By comparison, the stall force measured by laser tweezers during
DNA packaging in $\phi29$ was 57 pN \cite{smith}.  We would tend to think that the number
103 is a lower estimate for the actual force, because $\Delta \phi$
is underestimated at 72 C
as explained
in Subsection \ref{sect8.3}.  If we divide this force
by the cross-sectional
area of the sheath to get a stress, we get about 0.5 MPa.  This
is quite low as a (maximum) transformation stress in a macroscopic
crystalline martensitic material.  However, the transformation
strain in T4 sheath is enormous, so, if we calculate the energy
density  of contraction based on these numbers, we get numbers
that are comparable to those measured in
the best shape memory materials, which themselves exhibit the
highest energy densities in any known actuator system
\cite{pk}.

The transformation can also be induced by applying a
pure axial moment.
This leads to the analogue of  (\ref{efeten1})-(\ref{efeten4}),
except using the loading device energy $-22 M \gamma$, and
gives the moment at transformation of
\beqs
M &=& M_{trans}  =  \
\frac{(\phi_c(\rho^e, \lambda^e) - \phi_c(\rho^c, \lambda^c))\lambda_c}
{ (\gamma^e - \gamma^c)} \nonumber \\
&=& \frac{(\phi_c(\rho_e, \lambda_e) - \phi_c(\rho_c, \lambda_c)))\lambda_c}
{\gamma_e - \gamma_c} +
{\rm O}\left(\max_{c,e}\frac{A_{c,\, e}}
{A_{c,\, e}C_{c,\, e} - B_{c,\, e}^2} M^2   \right). \label{efemom}
\eeqs
Neglecting the higher order terms in (\ref{efeten4}) and
(\ref{efemom}) we have the simple approximate relationship
between the forces and moment needed to cause transformation
in the sheath:
\beq
  f_{trans} \approx \frac {\gamma_e-\gamma_c}{\lambda_e - \lambda_c}
   M_{trans}  =  - \frac{1}{89\ {\text \AA}}\ M_{trans}.
\eeq

\subsection{Biomolecular epitaxy, patterning and devices}

In this section we explore some more speculative ideas.
T4 tail sheath is a kind of biomolecular actuator, and
one could imagine that it could function as part of a man-made
machine \cite{kb_rdj_05} that could interact in an intimate
way with biological organisms.  Tubes of
{\it polysheath} several microns long
 have been synthesized \cite{kboy_64},
\cite{to_69}; polysheath is similar to,
but not exactly the same as, contracted sheath.  Tubes of
extended sheath  have not been synthesized separate from the
the baseplate and tail tube,
and this is understandable in view of their higher free energy
and also the possible role of the baseplate in stabilizing
extended sheath via epitaxy.

\begin{figure}
\includegraphics[width=.9\columnwidth]{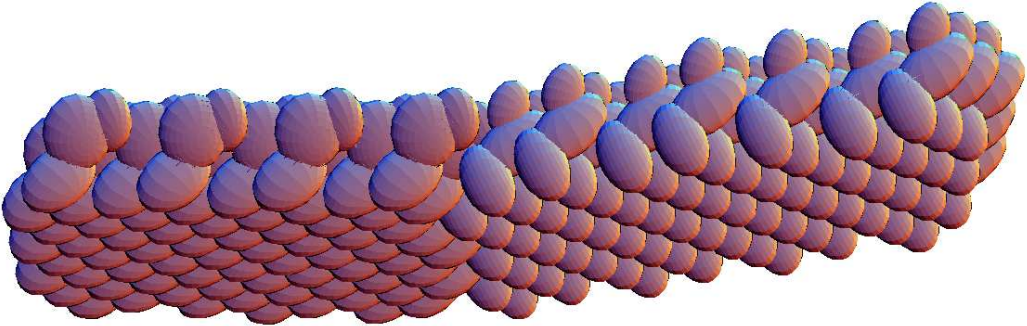}
\caption{Edge-on view, slightly below horizontal, of the
interface between contracted (left) and extended (right) sheath.}
\label{figure11}
\end{figure}

One basic interesting line of thought is to consider the
possibility of changing the heights of the energy wells.
As is true in a great many biological systems, hydrophobicity
plays a critical role, and this can be appreciated
in the present case by looking at Figure \ref{figure4}.
There, it is clear by inspection that extended sheath exposes
substantially more surface area to the surrounding solution
than contracted sheath,
and this qualitatively explains its higher free energy.  But it
also indicates that the relative free energies of extended
and contracted sheath are amenable to adjustment via manipulation
of the solution chemistry.  Systematic studies \cite{to_69}
of the effect
of solution chemistry on the breakup of parts of the virus
(capsid, neck, tail sheath, tail tube,  baseplate, tail fibers)
demonstrate sensitivity to solution chemistry.
Apparently, solutions
that cause contracted tail sheath to extend have not yet
been found. However, if by this means one could
exchange the heights of the wells, then tail sheath would
be like a shape memory material.  In a highly schematic way,
one could alter the solution so that extended sheath is
stable.  Then one could add an axial tensile force to the sheath.
Again manipulating the solution, one could return it to
phage-physiological conditions and, if the force was not too large
(i.e., below the value $f_{trans}$ of (\ref{efeten4})) then
the sheath would transform to the contracted form, while
doing work on the force.  This would be a machine that converts
chemical free energy of the solution to mechanical energy.
The very small cross-section of the sheath would allow it to
target a small region of a cell.
One could consider the possibility of vast arrays of these tubes.
In this regard, we note that ordered planar arrays of whole viruses
have been deposited on surfaces (not using epitaxy) by
Lee et al. \cite{lee}.

\begin{figure}
\includegraphics[width=.9\columnwidth]{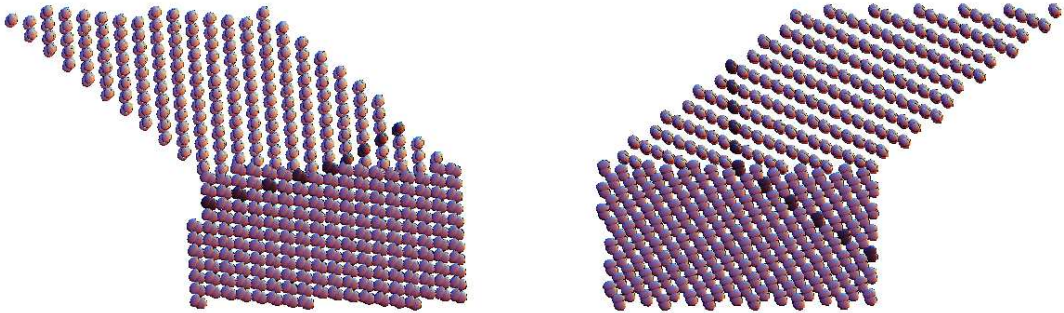}
\caption{The two compatible interfaces that separate
extended (above) from contracted (below) sheath.  In each
case the dark line of atoms was the main helix before unrolling
the sheath. }
\label{figure12}
\end{figure}

In the following discussion we allow rather drastic changes of
the sheath, but we enforce the constraints (\ref{con1})
and (\ref{con2}), these being in our view fundamental to its
behavior.  T4 sheath is in the shape of a cylinder, but it
is interesting to think about the possibility of slitting along
a generator and unrolling it.  We first note that it is possible
to do this without violating the constraints.  Secondly, extended
tail sheath exhibits an epitaxial relation to the baseplate, and
this likely plays a role in self-assembly of the sheath during
formation and subsequent stabilization.  Thus we suggest the
possibility of growing films of tail sheath epitaxially.
In general, epitaxial growth is aided by a substrate with the
same lattice parameters as the sheath, that is also chemically
compatible with the sheath.  The most likely possibility is
to grow the lower free energy contracted form (see the left
of Figure \ref{figure12} which shows the epitaxial surface).
It is interesting to note that epitaxial growth of protein
sheets could possibly take advantage
of the shapes of molecules and
the presence of functional groups, in addition to the
matching of lattice parameters and use of surface
chemistry, the latter principles familiar from the epitaxial growth
of semiconductors.

While it does not violate the constraints, unrolling is a
pretty drastic distortion, so one can expect some deviation
of the lattice vectors from the values
$\tvec_0^{c, e}, \tvec^{c, e}$ of Section \ref{sect7}.
Nevertheless, in the analysis below, we do use those values,
together with the orientations $\Rvec_{1,1}^{c,e}$.
Without loss of generality (using frame-indifference)
we first rotate these vectors
into the 1,2-plane (we do not relabel the resulting vectors)
and we apply the same rotations to the orientations.
We plot
the sheet as $i \tvec_0 + j \tvec$ where $i$ and $j$ are
integers.

The transformation matrix $\Gvec$ that maps $\tvec_0^c, \tvec^c$
into $\tvec_0^e, \tvec^e$  is the matrix
\beq
\left( \begin{array}{cc} 0.053 & -1.088  \\
0.999   & 1.543  \end{array} \right).
\eeq
By direct calculation $\Gvec^T \Gvec$ has eigenvalues
$2.06, 0.567$.  As is known from the theory of martensitic
transformations in sheets\footnote{see, e.g., \cite{kb_rdj_02}
which shows how to calculate these lines and the
corresponding deformations.  The presence of these
two interfaces was first noticed by Olson and Hartman
\cite{O1982}}, the fact that
these values straddle 1 (i.e., $2.06 > 1 >  0.567$) means that
there are exactly two interfaces on the sheet where extended
and contracted sheet meet compatibly.  These are pictured
in Figure \ref{figure12}.  We have rotated the sheets suitably
so that the interfaces are horizontal. The original orientation
can be inferred from the dark lines of atoms, which correspond
to what was the main helix (cf. Figure \ref{figure5}) before
unrolling. A cross section of the interface on the right is
seen in Figure \ref{figure11}.

\begin{figure}
\includegraphics[width=.9\columnwidth]{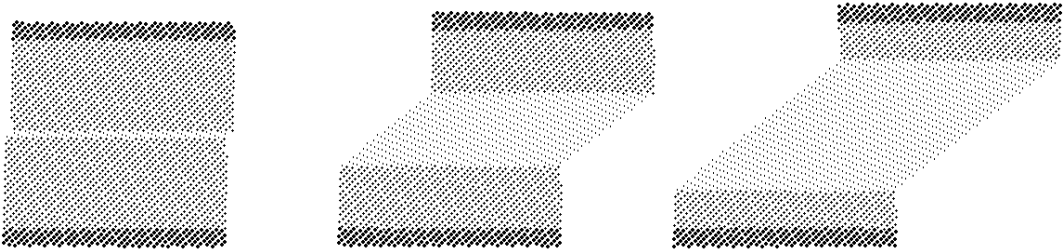}
\caption{Shear-inducing the transformation from contracted
to extended sheath.  Drawn with the lattice parameters of
extended and contracted sheath but with molecules represented
by dots.  The molecules are released from the substrate
except for the two dark strips at the top and bottom of
the sheet. }
 \label{figure13}
\end{figure}

Note that on the left of Figure \ref{figure12} the dark
line of atoms is approximately in the direction of the
interface.  This reflects the constraint, which embodies
the idea that the two phases are approximately equally
stretched along this line.  The reason that this line
does not exactly coincide with the interface is related
to the use of the effective radius, rather than the actual
radius, in (\ref{con1}).  As mentioned above, the values of
lattice parameters are likely to change a bit with ``unrolling'',
leading to interfaces that differ somewhat from those shown in
Figure \ref{figure12}.

From Figure \ref{figure12} one can imagine the possibility
of stress-inducing the transformation by shear as shown
in Figure \ref{figure13}.  This would provide a direct measure
of the relative heights of the energy wells and therefore of
the contraction force. Ideally, one could begin with an
epitaxially grown sheet, as discussed above, and release
the film from the substrate on the medium gray region
of Figure \ref{figure13}; some of the techniques
developed in the microactuator community \cite{dong}(such as
backside etching)
for patterning and releasing single crystal films could
be relevant.  Then by applying shear and slight extension as
shown in Figure \ref{figure13} the phase transformation could
be made to occur.
 Technically, the corners between phases
may introduce stress concentrations in such an experiment, but
this can be overcome by using a suitable indentor that induces
an appropriate out-of-plane deformation.  Once again, it would
be fascinating to bring chemistry into such an experiment by
altering the solution around the sheet.

\begin{appendix}
\section{Equations for the moments produced
by helical configurations}
 
This is a proof of (\ref{intequil})$_2$ and (\ref{momel}).
  Let $i \in \{ 2, \dots, N-1 \}$.
The terms of the energy (\ref{tfe}) that contain $\Rvec_i$ are
\beq
  \vphi(\Rvec_{i-1}^T(\yvec_{i} - \yvec_{i-1}),
  \Rvec_{i-1}^T \Rvec_i)   +
   \vphi(\Rvec_{i}^T(\yvec_{i+1} - \yvec_{i}),
  \Rvec_{i}^T \Rvec_{i+1})
\eeq
In this expression replace $\Rvec_i$ by
$(\Ivec +s \Wvec + \dots )\Rvec_i$, $\Wvec^T = -\Wvec$,
 differentiate with
respect to $s$ and evaluate at $s = 0$.  This gives
(\ref{intequil})$_2$ together with the formula
\beq
 (\mvec_{\ell,\ell+1})_j = \eps_{ijk}  \left(\Rvec_{\ell}
 \frac{\partial
 \vphi \left(\tvec_{\ell}, \Qvec_{\ell}\right)}{\partial \Qvec}\right)_{im}
 (\Rvec_{\ell+1})_{km}.
 \label{mome2}
\eeq
Components are with respect to the rectangular Cartesian
orthonormal basis used in the paper.
So, we need to show that  (\ref{mome2})
reduces to (\ref{momel}). In (\ref{mome2}) write $\Rvec_{\ell+1}
 = \Rvec_{\ell} \Qvec_{\ell}$ and define
 \beq
 \Svec_{\ell} = {\rm Skew} \left( \frac{\partial \vphi(\tvec_{\ell}, \Qvec_{\ell})}
 {\partial \Qvec} \Qvec_{\ell}^T \right),
 \eeq
where,  for any matrix $\Avec$, Skew$\Avec =
\frac{1}{2}(\Avec - \Avec^T)$. With this definition
 (\ref{mome2}) becomes
\beq
 (\mvec_{\ell,\ell+1})_j = \eps_{ijk}  \left(\Rvec_{\ell}
\Svec_{\ell}
 \Rvec_{\ell}^T \right)_{ik}.
 \label{mome3}
\eeq
Recalling the notation
(\ref{rots2}), (\ref{rots3}), we have
\beq
 \frac{\partial \vphi(\tvec_{\ell},\Qvec_{\ell})}{\partial w_j} =
 \eps_{ijk}\, (\Svec_{\ell})_{ik}.  \label{mome4}
\eeq
Multiply this by $\frac{1}{2}\eps_{sjt}$ and simplify to get,
\beq
(\Svec_{\ell})_{st} = \frac{1}{2}  \eps_{sjt}
 \frac{\partial \vphi(\tvec_{\ell},\Qvec_{\ell})}{\partial w_j}.
\eeq
Now use the identity $\eps_{jkl} R_{ij}  = \eps_{ipq} R_{pk} R_{ql}$
which holds in an orthonormal basis for any $\Rvec \in$ SO(3)
(i.e., invariance of the cross product under rotations) in the
component version of (\ref{mome3}):
\beqs
 (\mvec_{\ell,\ell+1})_j &=& \eps_{ijk}  \left(\Rvec_{\ell}
\Svec_{\ell}
 \Rvec_{\ell}^T \right)_{ik},    \nonumber \\
&=& - \eps_{jik} \, (\Rvec_{\ell})_{is} (\Svec_{\ell})_{st}
(\Rvec_{\ell})_{kt},          \nonumber \\
&=& - \eps_{lst} \, (\Rvec_{\ell})_{jl} (\Svec_{\ell})_{st},
 \nonumber \\
&=&   (\Rvec_{\ell})_{jl} \, \eps_{slt} \, (\Svec_{\ell})_{st},
 \nonumber \\
 &=&   (\Rvec_{\ell})_{jl} \, \frac{\partial \vphi(\tvec_{\ell},
 \Qvec_{\ell})}{\partial w_l};
   \label{mome5}
\eeqs
the last step follows from (\ref{mome4}).  This is  (\ref{momel}).

\section{Approximation of electron density maps}

Many arguments of this paper relied on the approximate shape
and orientation of the molecules.
In order to have a reasonable but fairly simple
representation of the molecules
of tail sheath, we approximated the electron density maps
of  Leiman et al. (\cite{leiman}; we are grateful to Petr Leiman
for providing prepublication data from high resolution cryo-electron
micrographs of extended sheath).
Information about how the positions and orientations were
extracted from the representations is described at the
end of Section \ref{sect7.1}.

The maps themselves showed clearly the presence of domains.
These have the general appearance of nestled ellipsoids,
so we approximated them by overlapping ellipsoids.
 We did this by partitioning the data by domain
(the colored regions of Figure \ref{figure14}).  Then we
computed the total electronic charge and center of mass
of each domain. Using charge neutrality we made mass density
proportional to electronic density.   This can have errors
arising mainly from the presence of H atoms, but in fact
the total charge of a domain (or
molecule) in contracted vs. extended state differed by
more than this error, so the quality of the data did not
justify a more detailed analysis.

With the center of mass
of the domain now fixed, we adjusted the principle axes of
the ellipsoids to match approximately the sectional
data.  Sections of the selected ellipsoids are shown
in Figure  \ref{figure14} superimposed on the data.
In this figure the concentric circles define the axis of
the tail sheath and the Z values indicate the slices of the
electron density map which were averaged.  Here Z is an
axial variable measured from a fixed (but arbitrary) reference.
Domains for both extended and contracted sheath are shown.
 The ellipses are
sections of the ellipsoid (for the enclosed domain)
at the corresponding average value of Z (i.e., for Z$ =$ 0 to 10
the slice through the ellipsoid was taken at Z $= $ 5).

\begin{figure}
\includegraphics[width=.9\columnwidth]{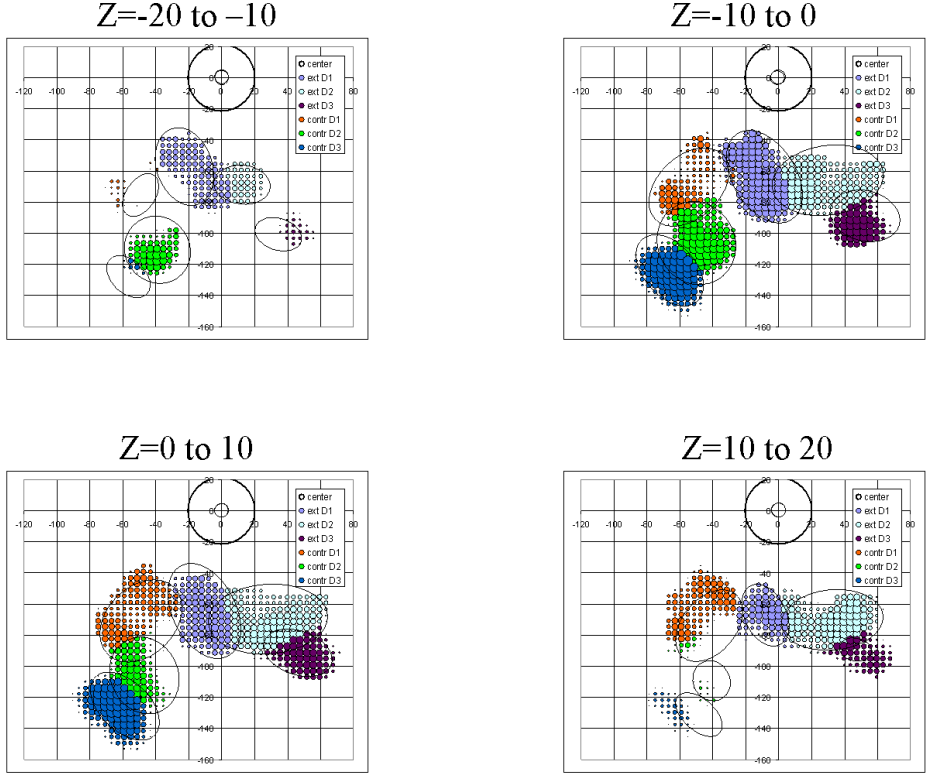}
\caption{Comparison of electron density maps and ellipsoidal
approximations.  See text. }
 \label{figure14}
\end{figure}

\end{appendix}

\begin{acknowledgments}
We thank Petr Leiman, Victor
Kostyuchenko and Michael Rossmann for the EM data and
helpful discussions, and we wish to acknowledge helpful
discussions with Gero Friesecke and John Maddocks.
This work was supported by NSF-NIRT
DMS-0304326 and by the U.S.~Army
High Performance Computing
Research Center under the auspices of the U.S.~Department of
the Army,
Army Research Laboratory cooperative agreement number
DAAD 191-01-2-0014.  The content does not necessarily reflect the
position or the
policy of the government, and no official endorsement should
be inferred.
\end{acknowledgments}
\bibliography{PRE.bib}
\end{document}